\documentclass[acmsmall]{acmart}

\usepackage{braket}

\usepackage{xcolor}
\definecolor{mmcolor}{rgb}{0.2,0.6,0.5}

\newcommand{\mm}[1]{\textcolor{black}{#1}}

\AtBeginDocument{%
   \providecommand\BibTeX{{%
    \normalfont B\kern-0.5em{\scshape i\kern-0.25em b}\kern-0.8em\TeX}}}

\setcopyright{acmcopyright}
\copyrightyear{2020}
\acmYear{2020}
\acmDOI{10.1145/1122445.1122456}

\acmJournal{TQC}
\acmVolume{37}
\acmNumber{4}
\acmArticle{111}
\acmMonth{8}

\begin{document}

\title{QASMBench: A Low-Level Quantum Benchmark Suite for NISQ Evaluation and Simulation}

\author{Ang Li}
\email{ang.li@pnnl.gov}
\orcid{0000-0003-3734-9137}
\author{Samuel Stein}
\email{samuel.stein@pnnl.gov}
\author{Sriram Krishnamoorthy}
\email{sriram@pnnl.gov}
\author{James Ang}
\email{ang@pnnl.gov}
\affiliation{%
  \institution{Pacific Northwest National Laboratory}
  \streetaddress{902 Battelle Blvd}
  \city{Richland}
  \state{WA}
  \country{USA}
  \postcode{99354}
}

%
%
%
%


\renewcommand{\shortauthors}{Li et al.}

\begin{abstract}
The rapid development of quantum computing (QC) in the NISQ era urgently demands a low-level benchmark suite and insightful evaluation metrics for characterizing the properties of prototype NISQ devices, the efficiency of QC programming compilers, schedulers and assemblers, and the capability of quantum system simulators in a classical computer. In this work, we fill this gap by proposing a low-level, easy-to-use benchmark suite called QASMBench based on the OpenQASM assembly representation. It consolidates commonly used quantum routines and kernels from a variety of domains including chemistry, simulation, linear algebra, searching, optimization, arithmetic, machine learning, fault tolerance, cryptography, etc., trading-off between generality and usability. To analyze these kernels in terms of NISQ device execution, in addition to circuit width and depth, we propose four circuit metrics including gate density, retention lifespan, measurement density, and entanglement variance, to extract more insights about the execution efficiency, the susceptibility to NISQ error, and the potential gain from machine-specific optimizations. \mm{Applications in QASMBench can be launched and verified on several NISQ platforms, including IBM-Q, Rigetti, IonQ and Quantinuum. For evaluation, we measure the execution fidelity of a subset of QASMBench applications on 12 IBM-Q machines through density matrix state tomography, which comprises 25K circuit evaluations. We also compare the fidelity of executions among the IBM-Q machines, the IonQ QPU and the Rigetti Aspen M-1 system.} QASMBench is released at: http://github.com/pnnl/QASMBench.

\end{abstract}



\maketitle


Quantum computing (QC) \cite{benioff1980computer, nielsen2002quantum} has been envisioned as one of the most promising computing paradigms beyond Moore's Law for tackling difficult computing challenges that are classically intractable arising from various domains like chemistry \cite{georgescu2014quantum, kandala2017hardware}, machine learning \cite{schuld2015introduction, biamonte2017quantum}, cryptography \cite{shor1999polynomial, gisin2002quantum}, linear algebra \cite{harrow2009quantum, clader2013preconditioned}, finance \cite{rebentrost2018quantum, woerner2019quantum}, recommendation systems \cite{kerenidis2016quantum}, physics simulation \cite{feynman1999simulating, divincenzo2000physical}, networking \cite{kimble2008quantum, munro2010quantum}, and so on. QC relies on fundamental quantum mechanisms such as superposition, entanglement, interference, tunneling, etc. for performing computation, in hopes of significant or even exponential speedups over existing algorithms in classical computers \cite{shor1999polynomial, coppersmith2002approximate}.

Despite holding great promise, QC in contemporary noisy-intermediate-scale-quantum (NISQ) \cite{preskill2018quantum} devices is still distant from outperforming classical computers regarding general problems. NISQ devices describe the near-term quantum platforms comprising 50 to hundreds of qubits, which is inadequate for full-scale error-correction but can already present promising results in a number of problems, and lay the foundations towards practical quantum demonstration \cite{preskill2018quantum}. \mm{The most recent public accessible NISQ device is IBM-Q Washington featuring 127 physical qubits.}

The QC landscape is evolving rapidly nowadays. From the software perspective, large number of QC software are developed in classical programming languages (e.g., Python \cite{cirq, steiger2018projectq}, C/C++ \cite{javadiabhari2014scaffcc}, JavaScript \cite{qiskit}, ML \cite{altenkirch2005functional}). A list of opensource QC projects, quantum algorithms, and quantum simulators can be found in \cite{qsoftware}, \cite{qalgorithmzoo}, and \cite{qsim}, respectively. From the hardware perspective, the development of QC testbeds have already proceeded to a stage where small working prototypes are available through cloud service, such as IBM-Q \cite{ibm}, Microsoft Azure \cite{azure}, \mm{Amazon Braket \cite{gonzalez2021cloud}}, and Rigetti \cite{rigetti}, showing encouraging results \cite{dumitrescu2018cloud, cincio2018learning}. Finally, from the application perspective, Google has announced the demonstration of quantum supremacy on the task of sampling the output of a pseudo-random quantum circuit, using a 53-qubit NISQ system \cite{arute2019quantum}. Several quantum machine learning frameworks (e.g., \cite{broughton2020tensorflow, qiskit}) and chemistry simulation packages have been released as well (e.g., \cite{mcclean2020openfermion, qiskit}).

\begin{figure}[!t]
\centering
\includegraphics[width=0.7\columnwidth]{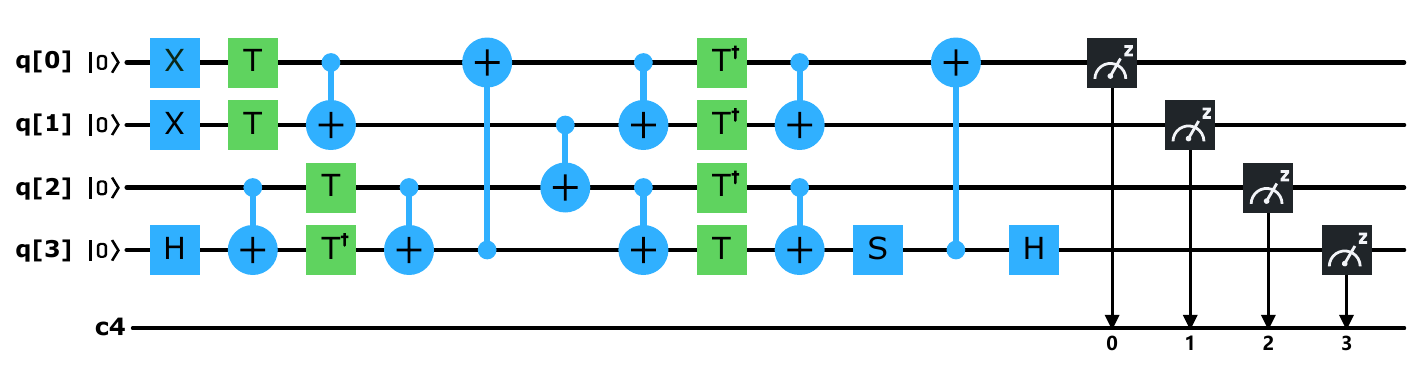} 
\caption{A quantum adder circuit with width=4 (4 qubits) and depth=23 (23 gates) followed by measurement.}
\label{fig:adder_n4}
\end{figure}

To close the gap between practical quantum applications and real quantum machines, contributions from the computer system and architecture community are undoubtedly critical in tacking several key challenges \cite{chong2018quantum}, including software and hardware verification \cite{tannu2019mitigating}, defining and perforating abstraction boundaries \cite{murali2019full}, managing parallelism and communication \cite{li2019tackling}, mapping and scheduling operations \cite{shi2019optimized}, elevating control complexity \cite{li2020towards}, hardware-specific optimizations \cite{tannu2019not, murali2019noise, smith2019quantum}, investigating and mitigating noise \cite{murali2020software, tannu2019ensemble}, etc. These massive effort, however, are currently evaluated and verified using randomly selected QC benchmarks \cite{li2019tackling, shi2019optimized, li2020towards, tannu2019not, murali2019noise, murali2020software, das2019case, tannu2019mitigating, tannu2019ensemble}, lacking the common ground for judicious analysis and fair comparison among each other. The communities thus urgently desire a low-level, easy-to-use QC benchmark suite to foster the software-hardware codesign effort for the purpose of validation and verification. This is particularly vital in the NISQ era, given the noise and technology limitation continuously to be the major challenges in the near feature. 

In this work, we propose a low-level benchmark suite called QASMBench based on the OpenQASM assembly-level intermediate representation (IR) \cite{cross2017openqasm}. It collects commonly used quantum algorithms and routines (e.g., the adder circuit in Figure~\ref{fig:adder_n4}) from a variety of distinct domains, including quantum chemistry, simulation, linear algebra, searching, optimization, arithmetic, machine learning, fault tolerance, cryptography, etc. The design of the benchmark suite trades-off between generality and usability, covering a wide spectrum regarding circuit depth (i.e., number of gates) and width (i.e., number of qubits). Additionally, to analyze and compare the benchmark applications, we propose four circuit evaluation metrics including gate density, retention lifespan, measurement density, and entanglement variance, to offer deeper insights about the execution efficiency, the susceptibility to system/readout error, and the potential impact from device-specific optimizations, when mapping to a NISQ device. \mm{Circuits from QASMBench can be directly uploaded and evaluated in IBM-Q machines as well as other platforms (e.g., Rigetti through the Qiskit interface, IonQ and Quantinuum through Microsoft Azure Quantum \cite{azure}). Additionally, opensource tools such as \emph{q-convert} and \emph{QCOR} allow the translation of QASMBench to other QC programming languages or representations.}

\mm{We evaluate QASMBench on 12 IBM-Q machines, and show the fidelity of circuit executions. The fidelity values are measured by calculating the Hellinger distance between the reconstructed density matrix of the resultant mixed state in a real quantum device through density matrix tomography, and the density matrix of the pure state by running the same circuit in a noiseless classical simulator. We observe that shallow circuits in general exhibit higher fidelity than deep circuits, likely due to less decay impact from a shorter execution time; whereas large quantum machines, such as the 127-qubit IBM-Q Washington, show reduced fidelity than small machines, despite using their best performing qubits. We also compare the fidelity of 4 QASMBench circuits among IBM-Q superconducting machines, IonQ's QPU trapped-ion machine, and Rigetti's Aspen M-1 superconducting machine, deriving interesting observations. Explanations are provided using the evaluation metrics proposed. Through such evaluation, we show how QASMBench and the 6 metrics can be used for evaluating contemporary and emerging NISQ platforms. }

This paper is organized as follows: In Section~1, we briefly describe NISQ and the landscape of OpenQASM. In Section~2, we introduce the QASMBench applications. \mm{In Section~3, we propose our characterization metrics. We characterize the QASMBench applications and evaluate them on IBM-Q and other NISQ devices in Section~4}. Finally, we summarize and draw the conclusion.

\section{Background: Quantum Computing in the NISQ Era}

We describe NISQ devices and the OpenQASM programming environment in this section.

\subsection{NISQ Devices}

\begin{figure}[!t]
\centering
\includegraphics[width=\columnwidth]{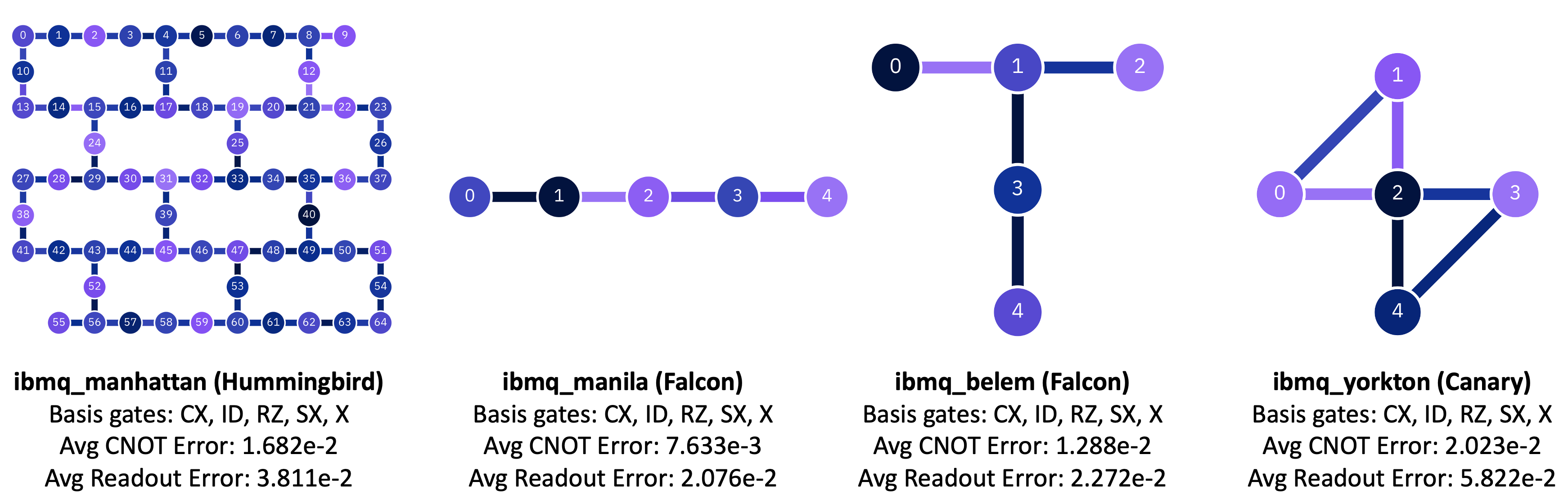} 
\caption{IBM-Q device topology. The color of nodes implies frequency (GHz) of the qubit or how fast a 1-qubit gate can be executed. The connection color implies the gate time in nanoseconds for 2-qubit gates such as \texttt{CX}. }
\label{fig:topo}
\end{figure}

NISQ devices describe the near-term quantum platforms incorporating 50 to less than a thousand qubits \cite{preskill2018quantum}. The qubits are currently built using a variety of different technologies, including superconducting qubits \cite{clarke2008superconducting, rigetti2012superconducting}, trapped-ions qubits \cite{cirac1995quantum, leibfried2003quantum}, spin qubits \cite{pla2012single, maurand2016cmos}, photonic qubits \cite{o2009photonic, aspuru2012photonic}, etc. To correctly run a QC circuit, the physical qubits have to stay coherent long enough for allowing the accomplishment of all the gate operations, thus imposing rigid constraints on allowable circuit depth \cite{corcoles2019challenges}. The \textbf{T1 coherence} time describes the time for a qubit to decay from the excited state $\ket{1}$ to the ground state $\ket{0}$. The \textbf{T2 coherence} time describes the time for a qubit to transit its state due to environment interaction \cite{tannu2019not}. Consequently, gate operations should be quickly and precisely performed on the qubits during the coherent window for maintaining the quantum states. The probability of involving error during the gate operation is known as \textbf{gate error}. The measurement operation, which reads-out classical output from a qubit by collapsing its final quantum state, is also critical for the correctness and efficiency of QC \cite{fu2017experimental}. The \textbf{readout error} describes the probability of inaccurate measurement of a qubit state \cite{zheng2020bayesian}. To precisely sample the output distribution with the existence of gate and read-out error, a quantum circuit is often executed many times (known as \textbf{shots} \cite{ibm} or repetitions \cite{cirq}).

\begin{table}[!t]
\centering\footnotesize
\caption{\mm{Basis gates for IBM-Q, Rigetti, IonQ and Quantinuum NISQ devices.}}
\begin{tabular}{|c|c|c|c|c|}
\hline
\textbf{Vendor} & \textbf{Technology} & \textbf{1-qubit basis gates} & \textbf{2-qubit basis gates} & \textbf{Reference}  \\ \hline
IBM-Q & Superconducting & \texttt{ID}, \texttt{RZ}, \texttt{SX}, \texttt{X} & \texttt{CX} & \cite{ibm} \\ \hline
Rigetti & Superconducting & \texttt{RX}, \texttt{RZ} & \texttt{CZ}, \texttt{XY} &  \cite{rigetti} \\ \hline
IonQ & Trapped-Ion & \texttt{GPI}, \texttt{GPI2}, \texttt{GZ} & \texttt{MS} &  \cite{ionq} \\ \hline
Quantinuum & Trapped-Ion & \texttt{RX}, \texttt{RZ} & \texttt{ZZ} & \cite{quantinuum}  \\ \hline
\end{tabular}
\label{tab:basis_gates}
\end{table}

\mm{NISQ devices are mainly featured by two attributes at the system level: \textbf{basis gates} and \textbf{topology}. The basis gate set, also known as \emph{instruction set architecture} (ISA) in classical computer architecture terminology, is hardware-defined. Upon execution, user quantum circuits are translated into an objective circuit only comprising basis gates. This process is called \textbf{quantum transpilation}. The basis gate sets for IBM-Q, Rigetti, IonQ and Quantinuum devices are listed in Table~\ref{tab:basis_gates}.}


\mm{Additionally, due to technical limitations, it is possible that only certain qubits of a NISQ device are physically connected. In other words, the physical qubits in a NISQ device may follow a certain \textbf{topology} (also known as coupling). Figure~\ref{fig:topo} illustrates the topology of 4 IBM-Q devices.} The topology limits the locations where two-qubit gates such as \texttt{CX} can be performed. If a \texttt{CX} is applied over two remote qubits, a series of \texttt{SWAP} gates are needed to relocate the two qubits to a connected tuple in the topology before the \texttt{CX} can be applied. For example, \texttt{CX(0,4)} in \emph{ibmq\_manila} in Figure~\ref{fig:topo} requires 4 extra \texttt{SWAP} gates. Additionally, more swaps may be needed to switch the qubit(s) back to their original position(s), adding extra overhead and error. \textbf{Qubit mapping} describes the process of mapping of the logical qubits of a quantum circuit to the physical qubits of a NISQ device, which significantly impacts QC accuracy and execution efficiency.

\subsection{OpenQASM Ecosystem}

OpenQASM (\emph{Open Quantum Assembly Language}) \cite{cross2017openqasm} is a low-level intermediate representation (IR) of quantum instructions, which is similar to traditional \emph{Hardware Description Language} (HDL) like \emph{Verilog} and \emph{VHDL}. OpenQASM is a unified low-level assembly language for IBM-Q quantum machines. These NISQ devices, being accessible through IBM-Q cloud or IBM-Q network \cite{ibm}, have been explored by many existing works \cite{tannu2019not, murali2020software, corcoles2019challenges, larose2019overview, murali2019full, murali2019noise, huang2020identification}. An OpenQASM code can be directly uploaded and verified in an IBM-Q machine or launched through Qiskit.

\begin{table}[!t]
\centering\scriptsize
\caption{OpenQASM gate definition (5 basic gates + 11 standard gates + 18 composition gates).}
\begin{tabular}{|c|l|c|l|c|l|}
\hline
\textbf{Gates} & \textbf{Meaning} & \textbf{Gates} & \textbf{Meaning} & \textbf{Gates} & \textbf{Meaning}  \\ \hline
\texttt{U3} & 3 parameter 2 pulse 1-qubit & \texttt{TDG} & conjugate of sqrt(S) &  \texttt{CRZ} & Controlled RZ rotation  \\ \hline
\texttt{U2} & 2 parameter 1 pulse 1-qubit  & \texttt{RX} & X-axis rotation    &  \texttt{CU1} & Controlled phase rotation   \\ \hline
\texttt{U1} & 1 parameter 0 pulse 1-qubit  & \texttt{RY} & Y-axis rotation  &  \texttt{CU3} & Controlled U3   \\ \hline
\texttt{CX} & Controlled-NOT & \texttt{RZ} & Z-axis rotation     &  \texttt{RXX} & 2-qubit XX rotation   \\ \hline
\texttt{ID} & Idle gate or identity & \texttt{CZ} & Controlled phase    & \texttt{RZZ} & 2-qubit ZZ rotation   \\ \hline
\texttt{X} & Pauli-X bit flip   & \texttt{CY} & Controlled Y   & \texttt{RCCX} & Relative-phase CXX    \\ \hline
\texttt{Y} & Pauli-Y bit and phase flip   & \texttt{SWAP} & Swap   & \texttt{RC3X} & Relative-phase 3-controlled X    \\ \hline
\texttt{Z} & Pauli-Z phase flip   &  \texttt{CH} & Controlled H   & \texttt{C3X} & 3-controlled X  \\ \hline
\texttt{H} & Hadamard   &  \texttt{CCX} & Toffoli  & \texttt{C3XSQRTX} & 3-controlled sqrt(X)   \\ \hline
\texttt{S} & sqrt(Z) phase  &  \texttt{CSWAP} & Fredkin   & \texttt{C4X} & 4-controlled X  \\ \hline
\texttt{SDG} & conjugate of sqrt(Z)  & \texttt{CRX} & Controlled RX rotation & &   \\ \hline
\texttt{T} & sqrt(S) phase   & \texttt{CRY} & Controlled RY rotation & & \\ \hline
\end{tabular}
\label{tab:gates}
\end{table}

Table~\ref{tab:gates} lists the types of gate that are defined in OpenQASM specification (i.e. the "\texttt{qelib1.inc}" header file) \cite{cross2017openqasm}. Within these gates, the first five, i.e., \texttt{U3}, \texttt{U2}, \texttt{U1}, \texttt{CX}, and \texttt{ID}, are \emph{basic gates} that are expected to be supported by the quantum backend. From \texttt{X} to \texttt{RZ} are \emph{standard gates} defined atomically in OpenQASM. These standard gates will be converted into basic gates during machine-specific assembling \& mapping phase before actual execution. The remaining gates from \texttt{CZ} to \texttt{C4X} are \emph{composition gates} that are constructed by standard gates. These gates are defined in \texttt{qelib1.inc} for the convenience of usage.

\begin{figure*}[!t]
\centering
\includegraphics[width=0.95\columnwidth]{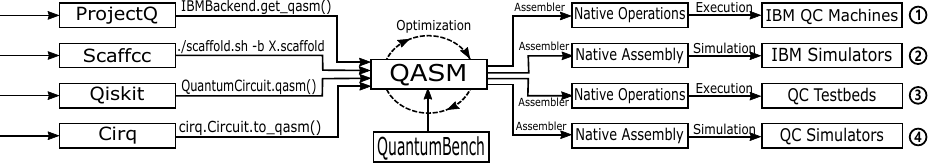} 
\caption{The OpenQASM toolchain.}
\label{fig:landscape}
\end{figure*}

OpenQASM is a low-level assembly language. It is executed "sequentially" without any loops, branches or jumps, making it very convenient for static analysis and simulating in a classical simulator. OpenQASM is widely supported. Several popular QC software frameworks such as Qiskit \cite{qiskit}, Cirq \cite{cirq}, Scaffold \cite{javadiabhari2014scaffcc}, ProjectQ \cite{steiger2018projectq} can dump the ready-to-execute circuit into an OpenQASM file, so it can be validated in the IBM-Q backends. Figure~\ref{fig:landscape} shows the landscape of OpenQASM. As can be seen, OpenQASM stands at the conjunction between quantum software and hardware, being critical for both platform-agnostic or platform-specific optimization, mapping, scheduling, evaluation, profiling, and simulation. An OpenQASM based benchmark suite can be useful for all these activities. In the following, we briefly discuss each of the front-end frameworks.

\vspace{4pt}\noindent\textbf{Qiskit:} The \emph{Quantum Information Software Kit} (Qiskit) \cite{qiskit} is a quantum software platform developed by IBM. Qiskit is mainly based on Python but is also available in JavaScript and Swift \cite{wilde2011swift}. The package comprises several tools, such as \emph{qiskit-aer} for simulation, \emph{qiskit-ignis} for hardware verification and noise addressing, and \emph{qiskit-aqua} for exemplary applications. \mm{At present, Qiskit is probably the most widely used quantum programming language. Both Rigetti \cite{rigetti} and Microsoft Azure \cite{azure} (offering IonQ and Quantinuum system access) provide Qiskit interface. OpenQASM can be easily dumped from a Qiskit program using "\emph{QuantumCircuit.qasm()}". Reversely, an OpenQASM circuit file can be loaded by Qiskit through "\emph{QuantumCircuit.from\_qasm\_str()}".}

\vspace{4pt}\noindent\textbf{Cirq:} Cirq \cite{cirq} is a QC software platform developed by Google. It is  based on Python. Despite claimed to be the interface for connecting to their 72-qubit Bristlecone quantum computer, this is not yet available to the general users. Cirq incorporates a local simulator for generic gates, and a \emph{Xmon-Simulator} for simulating the native gateset of Google's quantum computers. In addition, Cirq offers a function called "\emph{to\_qasm()}" in each circuit object so that a circuit can be dumped as an OpenQASM code that is runable on IBM-Q backends. Note, based on our experience, not all Cirq code can be translated to OpenQASM.

\vspace{4pt}\noindent\textbf{Scaffold:}  Scaffold \cite{javadiabhari2014scaffcc} is a quantum programming language embedded in the C/C++ programming language based on the LLVM compiler toolchain \cite{lattner2004llvm}. The major goal of Scaffold is to assist in developing quantum algorithms and advanced optimization, leveraging the existing LLVM compiling flow. By supporting advanced language structures such as loops, Scaffold can generate very complex OpenQASM code, such as the VQE examples in QASMBench. A Scaffold program can be compiled by \emph{Scaffcc} \cite{javadiabhari2014scaffcc} to native OpenQASM code using the "\emph{-b}" compiler option. 

\vspace{4pt}\noindent\textbf{ProjectQ:} ProjectQ is a quantum software platform developed by Steiger el al. from ETH Z\"urich \cite{steiger2018projectq}. Similar to Scaffold, ProjectQ does not have its own dedicated real quantum backend, but relies on classical simulation. However, it provides a way to generate OpenQASM code so that a ProjectQ program can be verified on an IBM testbed --- through "\texttt{IBMBackend.get\_qasm()}".

\begin{figure}[!t]
\centering
\includegraphics[width=0.6\columnwidth]{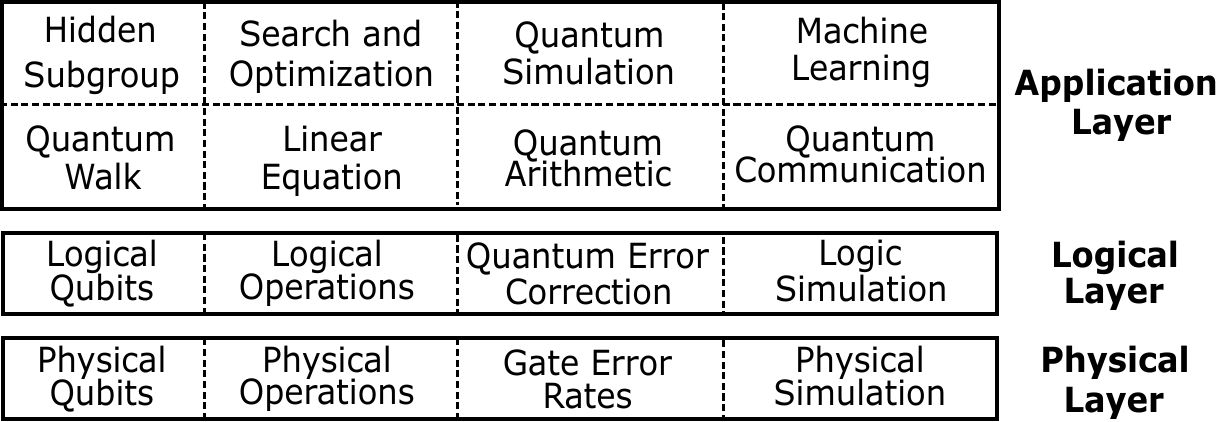} 
\caption{Quantum computing stack for the categorization of QASMBench. It is built based on \cite{montanaro2016quantum} by augmenting the categories of quantum arithmetic, quantum machine learning and quantum communication.}
\label{fig:stack}
\end{figure}

\section{QASMBench}
We introduce QASMBench in this section. QASMBench is an end-to-end package comprising a diverse variety of benchmark circuits, a system for evaluating the performance of a circuit, and characterizing circuit metrics for interpreting circuit characteristics both pre and post transpilation to real machines. Some of the benchmark routines are generated and reproduced from existing open-source QC software packages, including \cite{cross2017openqasm, javadiabhari2014scaffcc, sampaio96qc, michielsen2017benchmarking, mcclean2020openfermion, cirq, qiskit} through the approaches described in Figure~\ref{fig:landscape} whereas others are locally developed. 

\mm{Depending on the number of qubits used, QASMBench is partitioned into three categories:
\begin{itemize}
    \item \textbf{Small-scale}, with qubits ranging from 2 to 5. The purpose is to allow intensive measures such as density matrix tomography, with limited cost. Also, the present IonQ QPU and most IBM-Q devices with public access feature 5 qubits. The benchmarks in this category are listed in Table~\ref{tab:small}.
    \item \textbf{Medium-scale}, with qubits ranging from 6 to 15, for general benchmarking usage. Another historical reason is that when QASMBench was originally developed in 2020, the publicly accessible IBM-Q machine with the maximum number of qubits is \emph{ibmq\_melbourne}, which has 15 qubits. However, \emph{ibmq\_melbourne} is already retired now.
    \item \textbf{Large-scale}, contains benchmarks with more than 15 qubits, as listed in Table~\ref{tab:large}. The biggest circuit in QASMBench is \texttt{adder\_n127} that contains 3991 gates. This corresponds to the most recent IBM-Q machine -- \texttt{ibmq\_washington} which features 127 physical qubits.
\end{itemize}
 } 

For each item in Table~\ref{tab:small}, \ref{tab:medium} and \ref{tab:large}, we list its name, brief description, and the algorithm category it belongs to in the quantum stack (see Figure~\ref{fig:stack}). We show the number of qubits and gates utilized in the routines. The gate number here refers to \emph{standard} OpenQASM gates (not basic gates or composition gates) as discussed in Section~1.2. We also list the number of \texttt{CX} gates in the tables, which indicates a major source of delay and error. Each algorithm is briefly introduced under its respective category. However, certain routines appear in multiple categories, we only describe them in the first appearance. Last not the least, although QASMbench is developed based on OpenQASM, they can be converted to other representations such as Q\#, PyQuil, Cirq, etc. through the Javascript based \emph{q-convert} tool, which is available online: \url{http://quantum-circuit.com/qconvert}. \mm{We distribute QASMBench in other representations as well through a modified version of \emph{q-convert}.}

\subsection{Small-scale Benchmarks}

\begin{table*}[!t]
\centering\tiny
\caption{QASMBench Small-scale Benchmark.}
\resizebox{\textwidth}{!}{\begin{tabular}{|c|l|c|c|c|c|c|}
\hline
\textbf{Benchmark} & \textbf{Description} & \textbf{Algorithms} & \textbf{Qubits} & \textbf{Gates} & \textbf{CX} & \textbf{Ref} \\ \hline
\texttt{adder} & Quantum ripple-carry adder & Quantum Arithmetic & 4 & 23 & 10 &  \cite{javadiabhari2014scaffcc} \\ \hline
\texttt{basis\_change} & Transform the single-particle baseis of an linearly  connected electronic structure  & Quantum Simulation & 3 & 53 & 10 &  \cite{mcclean2020openfermion} \\ \hline
\texttt{basis\_trotter} & Implement Trotter steps for molecule LiH at equilibrium geometry  & Quantum Simulation & 4 & 1626  & 582 &  \cite{mcclean2020openfermion} \\ \hline
\texttt{bell\_state} & Bell State & Logical Operation & 2 & 3 & 1 &  \cite{weinfurter1994experimental} \\ \hline
\texttt{cat\_state} & Cat State & Logical Operation & 4 & 4 & 3 &  \cite{leibfried2005creation} \\ \hline
\texttt{deutsch} & Deutsch algorithm with 2 qubits for $f(x)=x$ & Hidden Subgroup & 2 & 5 & 1 & \cite{cross2017openqasm}  \\ \hline
\texttt{dnn} &  Quantum Deep Neural Network & Quantum Machine Learning & 2 & 268 & 84 & \cite{stein2021quantum}  \\ \hline
\texttt{fredkin\_n3} & Fredkin gate benchmark  & Logical Operation  & 3 & 19 & 9 & \cite{patel2016quantum}  \\ \hline
\texttt{qec\_dist3} & Error correction with distance 3 and 5 qubits  & Error Correction  & 5 & 114 & 49 & \cite{michielsen2017benchmarking}  \\ \hline
\texttt{grover} & Grover's algorithm   &  Search and Optimization & 2 & 16 & 2  & \cite{anakin}  \\ \hline
\texttt{hs4} & Hidden subgroup problem   &  Hidden Subgroup & 4 & 28 & 4  & \cite{javadiabhari2014scaffcc}  \\ \hline
\texttt{inverseqft} & Performs an exact inversion of quantum Fourier tranform   & Hidden Subgroup   & 4 & 8 & 0  & \cite{cross2017openqasm}  \\ \hline
\texttt{iSWAP} & An entangling swapping gate &  Logical Operation & 2 & 9 & 2  & \cite{javadiabhari2014scaffcc}  \\ \hline
\texttt{linearsolver} & Solver for a linear equation of one qubit & Linear Equation & 3 & 19 & 4 & \cite{quantumexamplebramdo2017} \\ \hline
\texttt{lpn} &  Learning parity with noise &  Machine Learning & 5 & 11 & 2 &\cite{sampaio96qc}  \\ \hline
\texttt{pea} &   Phase estimation algorithm  & Hidden Subgroup & 5 &  98 & 42  & \cite{cross2017openqasm}  \\ \hline
\texttt{qaoa} & Quantum approximate optimization algorithm  & Search and Optimization & 3 & 15 & 6 & \cite{qaoa3}  \\ \hline
\texttt{qec\_sm} &  Repetition code syndrome measurement  &  Error Correction & 5 & 5 &4  & \cite{cross2017openqasm}  \\ \hline
\texttt{qec\_en} & Quantum repetition code encoder   &  Error Correction & 5 & 25 & 10 & \cite{sampaio96qc}  \\ \hline
\texttt{qft} & Quantum Fourier transform  &  Hidden Subgroupe & 4 & 36 & 12 & \cite{cross2017openqasm}  \\ \hline
\texttt{qrng} & Quantum Random Number Generator  &  Quantum Arithmetic & 4 & 4 & 0 & \cite{cross2017openqasm}  \\ \hline
\texttt{quantumwalks} &  Quantum walks on graphs with up to 4 nodes & Quantum Walk & 2 & 11 & 3 & \cite{quantumwalk}  \\ \hline
\texttt{shor} & Shor's algorithm  & Hidden Subgroup  & 5 & 64 & 30 &  \cite{qiskit}  \\ \hline
\texttt{toffoli} & Toffoli gate  &  Logical Operation & 3 & 18 & 6 & \cite{javadiabhari2014scaffcc}  \\ \hline
\texttt{teleportation} & Quantum Teleportation  &  Quantum Communication & 3 & 8 & 2 & \cite{fedortchenko2016quantum}  \\ \hline
\texttt{jellium} & Variational ansatz for a Jellium Hamiltonian with a linear-swap network   & Quantum Simulation   & 4 & 54 & 16   & \cite{mcclean2020openfermion}  \\ \hline
\texttt{vqe\_uccsd} & Variational Quantum Eigensolver with UCCSD ansatz  &  Search and Optimization & 4 & 220 & 88 & \cite{grimsley2019trotterized}  \\ \hline
\texttt{wstate} &  W-state preparation and assessment  & Logical Operation &  3 &  30 & 9 & \cite{cross2017openqasm} \\ \hline

\end{tabular}}

\label{tab:small}
\end{table*}

\begin{table*}[!t]
\centering\scriptsize
\caption{QASMBench Medium-scale Benchmarks.}
\resizebox{\textwidth}{!}{\begin{tabular}{|c|l|c|c|c|c|c|}
\hline
\textbf{Benchmark} & \textbf{Description} & \textbf{Domain} & \textbf{Qubits} & \textbf{Gates} & \textbf{CX} & \textbf{Ref} \\ \hline
\texttt{dnn} &  Quantum Deep Neural Network & Quantum Machine Learning & 8 & 1200 & 384 & \cite{stein2021quantum}  \\ \hline
\texttt{adder} & Quantum ripple-carry adder & Quantum Arithmetic & 10 & 142 & 65 & \cite{cross2017openqasm} \\ \hline
\texttt{bb84} & A quantum key distribution circuit   & Quantum Communication & 8 & 27 & 0 & \cite{cirq}  \\ \hline
\texttt{bv} & Bernstein-Vazirani Algorithm & Hidden Subgroup & 14 & 41 & 13 &  \cite{cross2017openqasm} \\ \hline
\texttt{ising} & Ising model simulation via QC & Quantum Simulation & 10 & 480 & 90 &  \cite{javadiabhari2014scaffcc} \\ \hline
\texttt{multipler} & Quantum multipler   & Quantum Arithmetic & 15 & 574 & 246 & \cite{cirq}  \\ \hline
\texttt{multiply} & Performing 3$\times$5 in a quantum circuit & Quantum Arithmetic & 13 & 98 & 40 &  \cite{anakin} \\ \hline
\texttt{qaoa} & Quantum approximate optimization algorithm  & Search and Optimization & 6 & 270 & 54 & \cite{cirq}  \\ \hline
\texttt{qf21} & Using quantum phase estimation to factor the number 21 & Hidden Subgroup & 15 & 311  & 115 & \cite{anakin} \\ \hline
\texttt{qft} & Quantum Fourier transform  &  Hidden Subgroup &  15 & 540 & 210 & \cite{cross2017openqasm}  \\ \hline
\texttt{qpe} &   Quantum phase estimation algorithm  & Hidden Subgroup  & 9 & 123 & 43 & \cite{anakin}  \\ \hline
\texttt{sat} & Boolean satisfiability problem via QC &  Searching and Optimization & 11 & 679 & 252 & \cite{cross2017openqasm}  \\ \hline
\texttt{seca} & Shor's error correction algorithm for teleportation &  Error Correction & 11 & 216 & 84 & \cite{anakin}  \\ \hline
\texttt{simons} & Simon's algorithm & Hidden Subgroup & 6 & 44 & 14 & \cite{anakin}  \\ \hline
\texttt{vqe\_uccsd} & Variational quantum eigensolver with UCCSD  & Linear Equation  & 8 & 10808 & 5488 & \cite{javadiabhari2014scaffcc}  \\ \hline

\end{tabular}}
\label{tab:medium}
\end{table*}

\begin{table*}[!t]
\centering\scriptsize
\caption{\mm{QASMBench Large-scale Benchmarks.}}
\resizebox{\textwidth}{!}{\begin{tabular}{|c|l|c|c|c|c|c|}
\hline
\textbf{Benchmark} & \textbf{Description} & \textbf{Domain} & \textbf{Qubits} & \textbf{Gates} & \textbf{CX} & \textbf{Ref} \\ \hline
\texttt{adder} & Quantum ripple-carry adder & Quantum Arithmetic & 127 & 3991 & 910 &  \cite{javadiabhari2014scaffcc} \\ \hline
\texttt{bigadder} & Quantum ripple-carry adder & Quantum Arithmetic & 18 & 284 & 130 &  \cite{cross2017openqasm} \\ \hline
\texttt{bv} & Bernstein-Vazirani algorithm & Hidden Subgroup & 19 & 56 & 18 & \cite{cross2017openqasm} \\ \hline
\texttt{cat\_state} & Cat State & Logical Operation & 100 & 100 & 99 &  \cite{leibfried2005creation} \\ \hline
\texttt{cc} & Counterfeit coin finding problem via QC & Hidden Subgroup & 18 & 34  & 17 &  \cite{cross2017openqasm} \\ \hline
\texttt{ghz\_state} & GHZ State preparation and assessment & Logical Operation & 23 & 23 & 22 & \cite{greenberger1989going} \\ \hline
\texttt{ising} & Ising model simulation via QC & Quantum Simulation & 26 & 280 & 50 &  \cite{javadiabhari2014scaffcc} \\ \hline
\texttt{ising} & Ising model simulation via QC & Quantum Simulation & 60 & 654 & 118 &  \cite{javadiabhari2014scaffcc} \\ \hline
\texttt{multipler} & Quantum multipler   & Quantum Arithmetic & 25 & 3723 & 750 & \cite{cirq}  \\ \hline
\texttt{qft} & Quantum Fourier tranform & Hidden Subgroup & 20 & 970 & 380 & \cite{cross2017openqasm}  \\ \hline\
\texttt{qft} & Quantum Fourier tranform & Hidden Subgroup & 85 & 17935 & 7140 & \cite{cross2017openqasm}  \\ \hline
\texttt{square\_root} & Computing the square root of an number via amplitude amplification & Quantum Arithmetic & 18 & 4640 & 898 & \cite{javadiabhari2014scaffcc} \\ \hline
\texttt{swap\_test} &  SWAP Test algorithm implementation  & Logical Operation &  25 & 446 & 96 & \cite{kang2019implementation} \\ \hline
\texttt{wstate} &  W-state preparation and assessment  & Logical Operation &  27 & 446 & 96 & \cite{cross2017openqasm} \\ \hline

\end{tabular}}
\label{tab:large}
\end{table*}

\noindent\textbf{Adder}: The quantum adder uses each qubit as a classical bit, and builds a traditional adder through quantum gates. Here, we build the full-adder circuit using a basic component --- a 4-qubit ripple-carry adder \cite{vedral1996quantum, cuccaro2004new} with a carry-in and a carry-out qubit. Given the granularity of 4-qubits, a $m$-qubit adder would require $2m+m/4+1$ qubits in total where $m/4+1$ qubits are used for the carrying bits between the basic 4-qubit adder components. We included a 4-qubit adder (see Figure~\ref{fig:adder_n4}) in small-scale, a 10-qubit adder in medium-scale, \mm{and 18-/127-qubit adders in large-scale.} This benchmark is motivated by its scalability, easy-to-verify and usefulness in building quantum numeric units in the post-NISQ era.

\vspace{2pt}\noindent\textbf{Basis\_change}: The OpenQASM routine of this benchmark is generated from Google's Cirq based OpenFermion library \cite{mcclean2020openfermion} for manipulating fermionic systems towards chemistry simulation on quantum computers. It shows how to use a quantum circuit to transform the single-particle basis of an linearly connected electronic structure so as to realize exact evolution under a random Hermitian one-body fermionic operator. This is included as a quantum chemistry benchmark.

\vspace{2pt}\noindent\textbf{Basis\_trotter}: This routine is also generated from the Cirq based OpenFermion library \cite{mcclean2020openfermion} using the method described in Figure~\ref{fig:landscape}. It is designed to implement the Trotter \cite{lloyd1996universal, berry2007efficient} steps for an actual molecule \emph{LiH} at equilibrium geometry, so as to simulate basis molecular Hamiltonians taking the following form $H = \sum_{pq} T_{pq} a^\dagger_p a_q + \sum_{pqrs} V_{pqrs} a^\dagger_p a_q a^\dagger_r a_s$. The circuit uses 4 qubits but is quite deep, challenging the T1 and T2 decay of a NISQ device. It is included as another  chemistry benchmark. 

\vspace{2pt}\noindent\textbf{Bell State}: The Bell state describes the Bell Inequality Test. This algorithm implements the Bell experiment on a quantum computer. This is a fundamental quantum mechanics experiment, and consists primarily of local entanglement, and low SWAP cost no matter the topology.

\vspace{2pt}\noindent\textbf{Cat state}: The Cat-state is an imperative quantum sub routine, and forms the basis of quantum networking's EPR pairs \cite{haner2021distributed}. The circuit comprises entangling $n$ qubits, where the readout bitstring of the algorithm is homogeneous in 0 or 1 (i.e. $00...0$ or $11...1$). The bell state challenges primarily the entangling capabilities of a system.

\vspace{2pt}\noindent\textbf{Deutsch}: The Deutsch-Jozsa algorithm \cite{deutsch1992rapid} is a generalization of Deutsch's algorithm. The problem is defined as -- given an oracle function $f\{0,1\}^{n}\to\{0,1\}$ which takes an $n$ binary digit as input, and outputs either 0 or 1. The objective is to find whether the outputs are the same for all of the inputs (say constant), or are one value (i.e., 0 or 1) for half of the inputs and the other (i.e., 1 or 0) for the remaining half (say balanced). Here, if the $2^n$ possible inputs are encoded into an n-qubit register, and if the oracle function can be expressed as gate operations in a black-box, the quantum computer can get the answer with only one-time evaluation to the oracle function whereas a deterministic classical computer would require $2^{n-1}+1$ times evaluations. This algorithm is among the first examples to demonstrate that a quantum algorithm can achieve exponentially speedup over any possible deterministic classical algorithms. It is included as a baseline circuit.

\vspace{2pt}\noindent\textbf{DNN}: Quantum deep neural networks \cite{stein2021quantum} is a quantum adaption of the classical neural network where variational quantum circuits with a layered-approach are parameterized by a bank of $\theta$ values. QNN consists of gates such as \texttt{RY($\theta$)}, \texttt{CX} and \texttt{CRY($\theta$)} for encoding classical data and weight information as well as introducing entanglement. Tasks such as classification are commonly used in evaluating these models where an attainable objective function is described alongside quantum gate differentiation, where the circuit is optimized. The domain of Quantum Machine Learning (QML) and QNN have drawn substantial attention recently \cite{broughton2020tensorflow, stein2020qugan} as the hope for a quantum advantage in machine learning, thereby motivating the inclusion of the quantum DNN benchmark.


\vspace{2pt}\noindent\textbf{Fredkin}: The Fredkin operation in Quantum Computing forms the foundation for the SWAP test, an important machine learning operation \cite{stein2020qugan}. Fredkin gates challenge topologies by interacting three qubits with each other.

\vspace{2pt}\noindent\textbf{QEC}: Quantum error correction (QEC) is to introduce redundancy into the original circuit by using additional physical qubits so that the state of the original circuit is effectively protected against noise. The surface code is often considered to be among the most promising QEC approaches as it can cope with error rates around $10^{-2}$ \cite{fowler2012surface}. The \emph{qec\_dist3} benchmark performing a 3-distance 5-qubit surface code which is developed based on \cite{michielsen2017benchmarking}. It is the smallest code that can correct arbitrary single-qubit error. QASMBench includes other QEC benchmarks, such as \textbf{qec\_en} and \textbf{qec\_sm} for encoder and syndrome measurement of repetition code, and \textbf{seca} for Shor's QEC for teleportation. QEC is perhaps the most important algorithm for NISQ and the success of QC in the long run, hence we include four QEC benchmarks in QASMBench.

\vspace{2pt}\noindent\textbf{Grover:}  Grover's algorithm \cite{grover1996fast} is a quantum algorithm for searching in a database, offering quadratic speedup over classical searching methods. It takes a black-box oracle realizing the function: $f(x)=1$ if $x=y$, $f(x)=0$ if  $x\ne y$, and find $y$ within a randomly ordered sequence of $N$ items using $O(\sqrt{N})$ operations and $O(N\log{N})$ gates with probability $p \ge 2/3$. Grover's algorithm is appealing in that it determines with high probability the unique input given the output is pre-known. Grover's algorithm is one of the fundamental quantum algorithms considering the importance of searching in the big-data era.  

\vspace{2pt}\noindent\textbf{HS4:}  Hidden subgroup problems \cite{javadiabhari2014scaffcc} is a quantum problem capturing problems such as factoring and graph isomorphism. The algorithm is important in the theory of Shor's algorithm, and hence plays an important role in the evolution of benchmarking quantum computers.

\vspace{2pt}\noindent\textbf{QFT \& InverseQFT}: The Quantum Fourier Transform (QFT) \cite{coppersmith2002approximate} (and its inverse) applies (inverse) Fourier transformation to the wave function amplitudes. It is a linear transform over the states of qubits, which is the quantum analogue of the discrete (inverse) Fourier transform. QFT is a basic component of many well-known quantum algorithms, including Shor's algorithm, quantum phase estimation, hidden subgroup problem, etc. Besides, it has some interesting features such as zero entanglement variance, as well be discussed later.

\vspace{2pt}\noindent\textbf{iSWAP}: The iSWAP operation is a challenging dual qubit operation comprising multiple CNOT operations as well as single qubit operations. It is well suited to challenge two local qubits on their gate performance.

\vspace{2pt}\noindent\textbf{LinearSolver}: This benchmark realizes the HHL solver \cite{harrow2009quantum} for solving a linear equation of 1-qubit. The HHL algorithm estimates the outcome of a scalar measurement on the solution vector to a sparse linear equation system. We originally attempted to dump the multi-qubits HHL routine from Cirq \cite{cirq}, but encountered error in generating the OpenQASM code due to the mismatch between Google's native quantum IR and OpenQASM (e.g., missing the \texttt{CCCRy} representation). HHL is a critical quantum algorithm given the importance of linear algebra in scientific computation and machine learning.

\vspace{2pt}\noindent\textbf{LPN}: The LPN (\emph{learning parity with noise}) benchmark is a machine learning problem of learning a hidden parity function defined by the unknown binary string in the presence of noise. This application is very promising for NISQ devices because (i) it is computationally intractable for classical approaches \cite{park2018noise}; (ii) it actually leverages the noise presented in the NISQ devices \cite{cross2015quantum}, which is usually seen as a hurdle in other applications.

\vspace{2pt}\noindent\textbf{Pea}: Quantum \emph{phase-estimation} (QPE) \cite{nielsen2002quantum} is an algorithm to compute the eigenvalue $e^{2\pi i \theta}$ of a unitary operator operating on $m$ qubits with the eigenvector $\ket{\psi}$ such that $U\ket{\psi}=e^{2\pi i \theta}\ket{\psi}$, $0\le\theta<1$. QPE is another fundamental quantum algorithm being the key component of Shor's method. This \emph{Pea} benchmark circuit describes a 2-qubit system including a read-out ancilla bit, and a physical system bit \cite{dobvsivcek2007arbitrary}, showcasing the property of QPE.

\vspace{2pt}\noindent\textbf{QAOA}: The quantum approximate optimization algorithm (QAOA) \cite{farhi2014quantum, zhou2018quantum} is another variational quantum algorithm that is particularly interesting to NISQ devices, given its ability to tolerant certain degree of noise through the unique quantum-classic hybrid algorithm design. QAOA is designed to solve combinatorial optimization problems \cite{zheng2020bayesian}, such as in graph analytics. In QAOA, a quantum subroutine is embedded in a classical search loop. The initial quantum state is prepared according to a set of variational parameters. These parameters are adjusted based on the measurement in the previous iteration. The QAOA benchmark in QASMBench represents the quantum workload for a sample iteration.

\vspace{2pt}\noindent\textbf{QuantumWalks}:
Quantum walks \cite{aharonov2001quantum, venegas2012quantum} is the quantum analogue of the classical random-walk algorithm, which has been adopted for many problems including graph isomorphism \cite{gamble2010two} and ranking nodes in a network \cite{paparo2012google}. Both discrete and continuous-time quantum-walk algorithms have been formulated by existing works \cite{aharonov2001quantum, farhi1998quantum}. The quantum-walk benchmark here represents the workload for a single step of the walks \cite{quantumwalk}. 

\vspace{2pt}\noindent\textbf{QEC SM \& EN}: Quantum Error correction codes are examples of current approaches to quantum error correction. These circuits form the foundation for Quantum Error Correction, and can be used to gain intuition as to the processors' performance for error correction.

\vspace{2pt}\noindent\textbf{QRNG}: Quantum random number generator is an algorithm for generating completely random numbers. It is known that no absolutely random random number generator exists, and is a relatively simple quantum circuit with no entangelement. This routine, although simple, can be very useful for real world applications.

\vspace{2pt}\noindent\textbf{Shor}: Shor's algorithm \cite{shor1994algorithms} is a famous QC algorithm for integer factorization in polynomial-time. It demonstrates theoretical exponential speedups over the classical algorithms, posing substantial threat and concern over the widely-used public-key cryptography systems like RSA. Given its importance in quantum cryptography, we include a 5-qubit sample here in QASMBench. We also include another example --- \textbf{qf21} for showcasing the process of factoring the integer of 21.

\vspace{2pt}\noindent\textbf{Toffoli}: Similar to the Fredkin gate, the Toffoli gate \cite{toffoli1980reversible} or \texttt{CCX}) gate is another universal gate. It refers to a circuit with three inputs and three outputs that inverts the third qubit if the first two qubits are both 1, otherwise all bits stay unchanged. We include the 3-qubit Toffoli gate as a baseline benchmark in QASMBench.

\vspace{2pt}\noindent\textbf{Teleportation}: The teleportation routine is a quantum communication channel connecting two qubits, communicating through an ancilla third. Quantum teleportation challenges a processors' ability to communicate information between an ancilla qubit.

\vspace{2pt}\noindent\textbf{VQE}: The variational quantum algorithm is to optimize a parameterized quantum circuit ansatz applied to some initial state for minimizing a cost function defined according to the output state \cite{mcclean2016theory, jones2019variational}. When applied in quantum simulation (i.e., simulating physical phenomena using QC rather than simulating QC in a classical computer), the goal of the algorithm is often to prepare ground states. The cost function is often the expectation value of a Hamiltonian. If the initial state is $\ket{\psi}$, the Hamiltonian is $H$, the ansatz is $U(\vec{\theta})$ where $\vec{\theta}$ is the variational parameter, then the cost function is $E(\vec \theta) = \bra{\psi} U^\dagger(\vec{\theta}) H U(\vec{\theta}) \ket{\psi}$. The \emph{Jellium} benchmark creates a variational ansatz for a Jellium Hamiltonian via a linear-swap network. We also include two \emph{Unitary Coupled Cluster Single-Double} VQE ansatz \cite{whitfield2011simulation} for estimating the ground energy of molecules due to their ultra deep circuits.

\noindent\textbf{W\_state}: \emph{w\_state} is one of the basic states exhibiting properties not observed in a classical system. The \emph{w\_state} \cite{dur2000three} is a special type of entangled quantum state generalized by:
\begin{equation*}
    \ket{W}=\frac{1}{\sqrt{n}}(\ket{10\dots0}+\ket{01\dots0}+\dots+\ket{00\dots1})\ \text{for}\ n>2
\end{equation*}
It describes the condition that for an n-qubit system, no matter which single qubit is measured or lost, the remaining (n-1) qubits still remain entangled. This property makes it useful for improving the robustness of ensemble-based quantum memories \cite{fleischhauer2002quantum}. This routine implements the 3-qubit condition where the three qubits follow $\ket{W}=\frac{1}{\sqrt{3}}(\ket{001}+\ket{010}+\ket{100})$.

\subsection{Medium-scale Benchmarks}

\vspace{2pt}\noindent\textbf{BB84}: \emph{BB84} \cite{bennett2020quantum} is a quantum key distribution (QKD) protocol which is the first quantum cryptographic protocol \cite{branciard2005security} based on the non-cloning quantum laws for showcasing provably secure key generation. It relies on the fact that it is not possible to obtain information distinguishing two non-orthogonal states without disturbing the signal. Since QKD is a critical component for quantum communication and quantum networks, we include BB84 as a sample benchmark in QASMBench.

\vspace{2pt}\noindent\textbf{BV}: The Bernstein-Vazirani algorithm \cite{bernstein1997quantum} is an extension to the Deutsch-Josza algorithm \cite{deutsch1992rapid}. The problem is that, given a black-box function $f$, which takes in $n$ bits as input, and outputs the bitwise product of a string $s$ against the input. This is defined as $x,f(x)=s\times xmod(2)$ where the string $s$ is hidden. Classically, to determine the hidden string, we need $n$ evaluations ($100..0,0100..0,...,0000..1$). This algorithm can decide $s$ through a single query to the oracle on a quantum computer. Given it is another application that can showcase exponential quantum speedup, we include it in QASMBench.

\vspace{2pt}\noindent\textbf{Ising}: The Quantum Ising Model \cite{labuhn2016tunable, chakrabarti2008quantum} consists of an array of quantum spins arranged in a certain lattice. The spins, which are expressed in quantum operators, can only interact with their neighbors. The phase transition is due to quantum fluctuation introduced by the transverse field. Given the quantum annealing machines (e.g., D-Wave) basically resolving optimization problems that can be expressed in an Ising model, we include it as a benchmark routine in QASMBench.


\begin{figure}[!t]
\centering
\includegraphics[width=0.7\columnwidth]{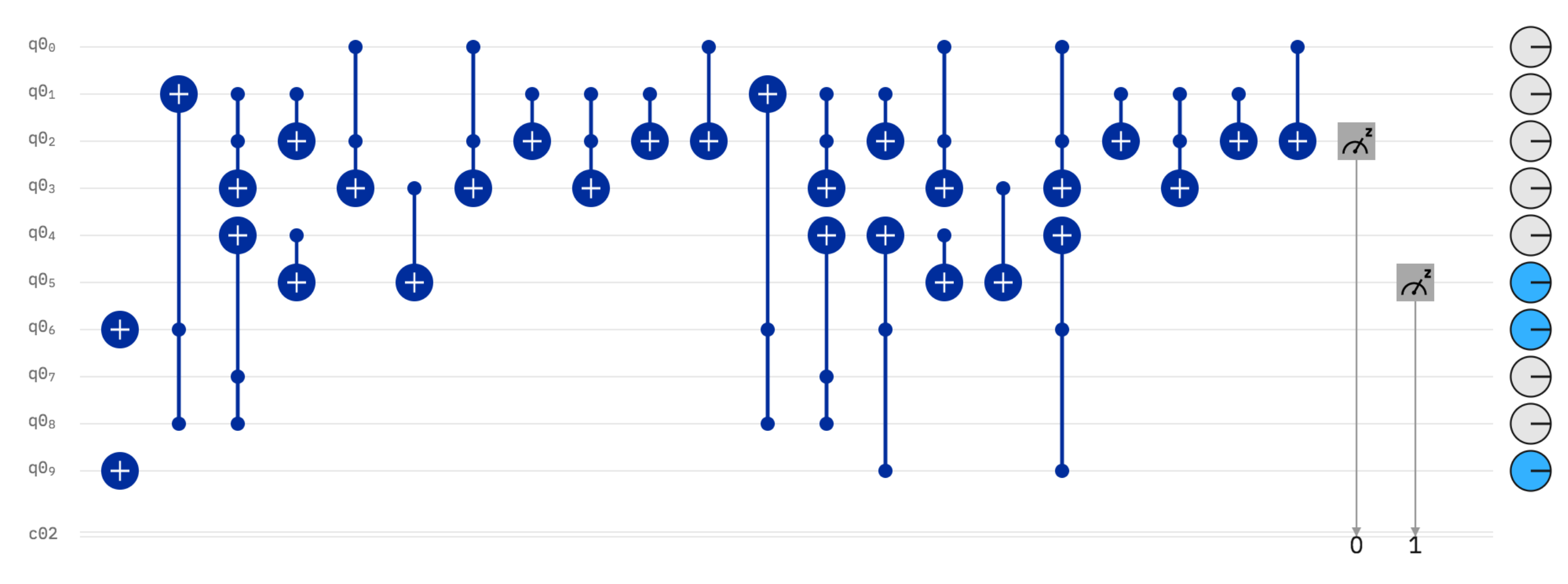} 
\caption{A 2-qubit multiplier. The execution phases include: (1) \emph{Preparation} (2) \emph{Execution} (3) \emph{Measurement}.}
\label{fig:multiplier}
\end{figure}

\vspace{2pt}\noindent\textbf{Multiplier}: Similar to the adder, the quantum multiplier also uses a qubit as a classical bit to construct a multiplier unit. Our implementation is similar to the one from Cirq \cite{cirq}, which requires $5n$ qubits for an $n$-qubit multiplier and internally calls the $n$-qubit adder by $n$ times for aggregating the partial results. The multiplier comprises the three phases: \emph{preparing operand}, \emph{multiply} and \emph{measure}. Figure~\ref{fig:multiplier} shows an example of a $2$-qubit multiplier using 10 qubits. Similarly to the adder, this is motivated by its simple scalability and easy verification.

\vspace{2pt}\noindent\textbf{SAT}: The \emph{Boolean Satisfiability} problem (SAT) determines whether there is an assignment of variables that satisfies a given Boolean function \cite{su2016quantum, barron2015classical} which is one of the first proven NP-complete problems. Since many optimization problem (e.g., in power-grid domain) can be attributed as a mixed-integer-linear-programming (MILP) problem, which can be formulated as a binary optimization problem \cite{chen2018toward} similar to SAT, we include SAT here as a benchmark in QASMBench.

\vspace{2pt}\noindent\textbf{Simon}: Simon's algorithm \cite{simon1997power} was among the first quantum algorithms to show exponential speedups over the best classical deterministic or probabilistic algorithm. It finds an unknown nonzero $s\in\{0,1\}^n$ that satisfies $f(x)=f(x\oplus s)$. We include it here as another benchmark showing quantum advantage. Besides, Simon's algorithm is more sensitive to noise in NISQ devices.

\subsection{Large-scale Benchmarks}

\vspace{2pt}\noindent\textbf{CC}: The \emph{counterfeit coin} problem is a mathematical puzzle that attempts to find all false coins from a given bunch of coins using a balance scale \cite{iwama2010quantum}. Let's say you have a set of N coins, and up to k of them are counterfeit and hence lighter. Using a scale, which returns the information of "balanced" or "tilted", we can deduce if any, or how many of the coins are counterfeit. The classical algorithm requires a time complexity of $O(k\log (k/N)$). The quantum algorithm offers an exponential speed-up, with a time complexity of $O(k^\frac{1}{4})$ \cite{iwama2012quantum}.

\vspace{2pt}\noindent\textbf{GHZ-state}: A \emph{Greenberger–Horne–Zeilinger} (GHZ) state \cite{greenberger1989going} is a particular type of entangled quantum state that involves at least three qubits: $\ket{GHZ}=\frac{1}{\sqrt{2}}(\ket{000}+\ket{111})$. Since it describes the superposition of all qubits being 0 with all of them being 1, it represents the maximally entangled quantum state. Consequently, characterizing GHZ state can provide very useful information about the fidelity of multi-qubit interactions, which is critical for constructing large-scale quantum computers. For an n-qubit system, GHZ state is generalized as:
\begin{equation*}
    \ket{GHZ}=\frac{1}{\sqrt{2}}(\ket{0}^{\otimes n}+\ket{1}^{\otimes n})\ \text{for}\ n>2
\end{equation*}
To construct a GHZ-state for an n-qubit system, from an all-zero states $\ket{\Psi}=\ket{0}^{\otimes n}$, after an \texttt{H} gate is applied on the first qubit, $\ket{\Psi}=\frac{1}{\sqrt{2}}(\ket{0}^{\otimes n}+\ket{1}\otimes\ket{0}^{\otimes n-1})$. We then apply a series of \texttt{CX} (or \texttt{CNOT}) gates from qubit 0 to qubit (n-1). Running on a full-fidelity quantum machine with 5 logic qubits, the measurement result should be 50\% $0^{\times n}$ and 50\% $1^{\times n}$. This application represents another basic quantum state which exploits across all the available qubits. It can be used to prove CNOT gate stability of a NISQ device.

\vspace{2pt}\noindent\textbf{Squarer\_root}: This routine relies on amplitude amplification \cite{grover1998quantum} to find the square root of an n-bit number using Grover's search technique. This is a complex application in QASMBench.

\begin{figure*}[!htb]
\minipage{0.33\columnwidth}
\includegraphics[width=1\columnwidth]{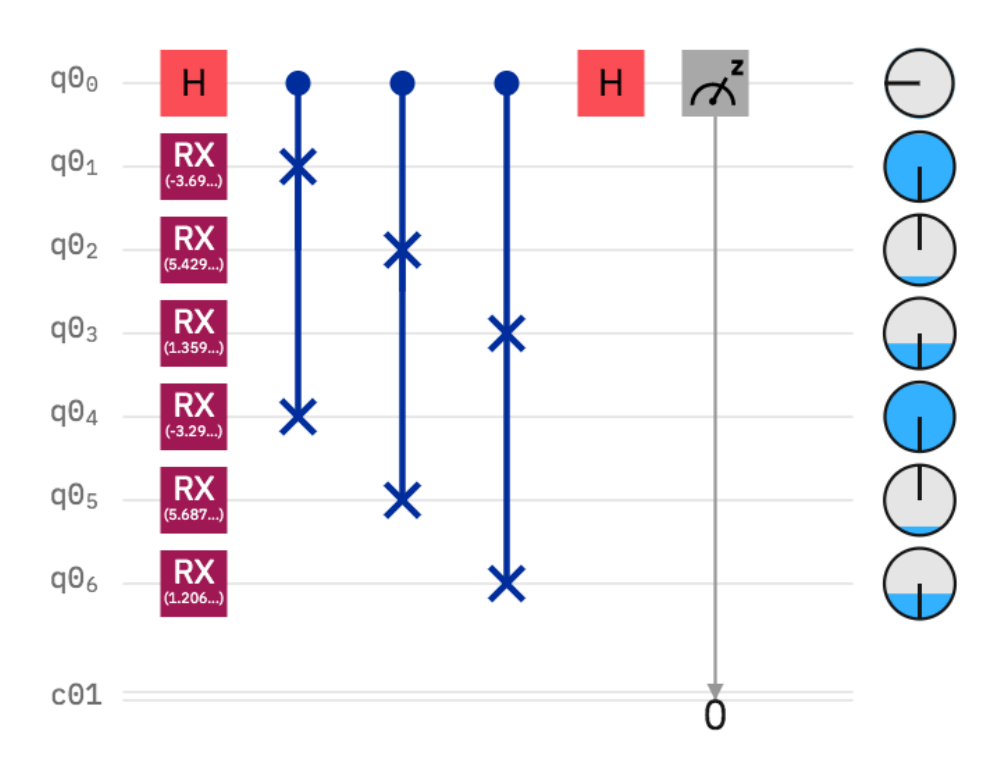} 
\caption{3-qubit swap\_test.}
\label{fig:swap_test}
\endminipage\hfill
\minipage{0.67\columnwidth}
\includegraphics[width=1\columnwidth]{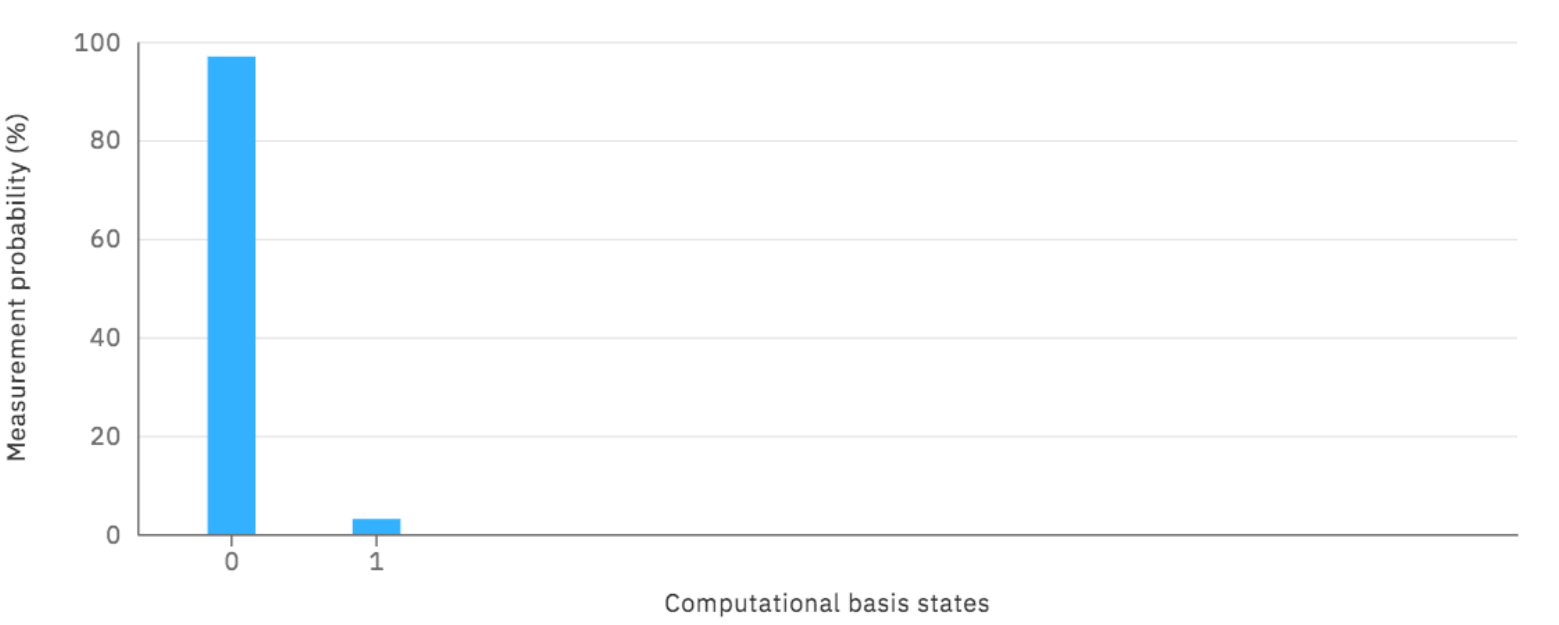}
\caption{Measurement result for a 3-qubit swap\_test circuit using 7 qubits.}
\label{fig:swap_test_res}
\endminipage
\end{figure*}

\vspace{4pt}\noindent\textbf{Swap-test}: 
The swap-test is a very convenient and economic way to measure the difference between two quantum states, or how two quantum ($\ket{\psi}$ and $\ket{\phi}$) states are divergent from each other. The swap-test can be very useful in quantum machine learning for estimating the prediction accuracy (e.g., distance of prediction from the tag). It uses an ancilla qubit for example qubit 0. The initial state of the system is $\ket{0,\psi,\phi}$. The ancilla qubit is firstly put in superposition via an \texttt{H} gate, as shown in Figure~\ref{fig:swap_test}, which converts the system state into $\frac{1}{\sqrt{2}}(\ket{0,\psi,\phi}+\ket{1,\psi,\phi})$. By applying the series of controlled-swap gate (or \texttt{CSWAP}, which swaps the states of two qubits only if the control qubit is 1, see Table~\ref{tab:gates}), the system state becomes $\frac{1}{\sqrt{2}}(\ket{0,\psi,\phi}+\ket{1,\phi,\psi})$. Then, by applying another \texttt{H} gate on qubit-0, the system transits to $\frac{1}{2}(\ket{0}(\ket{\psi,\phi}+\ket{\phi,\psi})+\ket{1}(\ket{\psi,\phi}-\ket{\phi,\psi}))$. When measuring qubit-0, the probability of obtaining 0 is $P(0)=\frac{1}{2}+\frac{1}{2}|\braket{\psi|\phi}|^2$. Therefore, if $\ket{\psi}$ and $\ket{\phi}$ are orthogonal, $|\braket{\psi|\phi}|^2=0$, the probability of obtaining 0 is 50\%; if $\ket{\psi}$ and $\ket{\phi}$ are equal, $|\braket{\psi|\phi}|^2=1$, the probability of obtaining 0 is 100\%. The more the two states are similar, the higher the probability of obtaining 0 approaches 1. Note that the swap-test needs sufficient number of repetitions to achieve an accurate sampling of the probability of the ancilla qubit. The measurement probability in Figure~\ref{fig:swap_test_res} shows that the two 3-qubit states in Figure~\ref{fig:swap_test} are similar to each other with $P(0)\approx1$. The SWAP test is a crucial algorithm in many Quantum processes, and is easily scaled out to any application size with substantial entanglement, hence is an ideal NISQ evaluation benchmark.

\section{QASMBench Characterization Metrics}

We propose a set of quantum circuit based evaluation metrics representing various features of a quantum application. These metrics are designed such that through them certain estimation can be performed on executing a particular circuit over a particular NISQ device, practicing the software-hardware co-design paradigm. The metrics serve as useful indicators on how a quantum circuit can stress a NISQ hardware device. The scripts for profiling these metrics are provided along with the QASMBench code. 

\subsection{Circuit Width}

Circuit width is defined as the number of qubits that enter the superposition state at least once within an application's lifespan. Qubits that are measured in the interim of a circuit and re-enter superposition are only counted as one qubit towards the circuit width. Circuit width dictates the spatial capacity required for a quantum device in order to run the quantum circuit. \mm{Circuit Width is mathematically defined in Equation~\ref{eqn:circuit_width} where $n_{{q_\text{active}}}$ is the number of qubits that are in-use or demanded by the circuit.}
\begin{equation}
    \text{Circuit Width} = n_{{q_\text{active}}}
    \label{eqn:circuit_width}
\end{equation}

\subsection{Circuit Depth}
Circuit depth is defined as the minimum time-evolution steps required to complete a quantum application. Time evolution is the process of completing all gates defined at time $t=t_j$, and once these are completed, the circuit moves onto time $t=t_{j+1}$, where the following gates are to be processed. To keep generality and avoid the impact from low-level optimization, circuit depth here is calculated solely using standard OpenQASM gates (see Section~1.2). That is to say, the OpenQASM code is decomposed to standard gates before counting the accurate time evolution steps required. As a concrete example, an \texttt{X} gate is a standard gate and hence requires 1 time-step, whereas a \texttt{CRZ} gate requires 4 standard gates to complete, thereby requiring 4 timesteps. 

Circuit depth can be computed by decomposing OpenQASM code into a $n_q \times t$ matrix $Q$, where $Q_{q_i,t_j}$ is the time-evolution steps to complete the gate on qubit $i$ at time $j$. The sum of the maximum time in each column is then equal to the minimum time required for a quantum application This is described in Equation~\ref{eqn:circuit_depth}: 
\begin{equation}
    \text{Circuit Depth} = \sum_{j=1}^{t} \max_{0\le i<\text{width}}(Q_{q_i,t_j})
    \label{eqn:circuit_depth}
\end{equation}

\subsection{Gate Density}
Gate density, or operation density, describes the occupancy of gate slots along the time-evolution steps of a quantum circuit. As certain qubits might need to wait for other qubits in the time evolution (i.e, gate dependency), they remain idle by executing the \texttt{ID} gate (see Table~\ref{tab:gates}). Consequently, if a gate slot is empty due to dependency, it implies a lower occupancy for the quantum hardware. This is similar to a classical processor, where data dependency introduces pipeline bubbles and reduced occupancy. We propose Gate Density to measure the likely occupancy of a circuit when mapping to a quantum hardware, as formulated in Equation~\ref{eqn:gate_density}. Note, the gate density is also calculated after decomposing down to the OpenQASM standard gates.
\begin{equation}
    \text{Gate Density} = \frac{G_\text{1-qubit}+2\times G_\text{2-qubit}}{\text{Circuit Depth}\times\text{Circuit Width}}
    \label{eqn:gate_density}
\end{equation}
where $G_{1-qubit}$ refers to the number of 1-qubit standard gates and $G_{2-qubit}$ refers to the number of 2-qubit standard gates (see Table~\ref{tab:gates}).

\subsection{Retention Lifespan}
Retention Lifespan describes the maximum lifespan of a qubit within a system, and is motivated by the T1 and T2 coherence time of a quantum device (see Section~1.1). A longer lifespan of a quantum system implies more decay to the ground state (T1) and state-transition due to environment noise (T2), thus is more susceptible to information loss. Therefore, we propose taking the qubit with the longest lifespan to determine the system's retention lifespan. Using this metric, one can estimate if a particular circuit can be executed in a NISQ device with high fidelity, given its T1/T2 coherence time. Note, all IBM-Q machines offer T1/T2 coherence time as status indicators for the hardware. As circuit depth can grow substantially, we introduce the log operator to shrink the scale. If $D_{i}$ is used to describe the lifespan or depth of the qubit $i$, Retention Lifespan, which measures the circuit size and sensitivity to quantum system error, is defined in Equation~\ref{eqn:retention_lifespan}:
\begin{equation}
    \text{Retention Lifespan} = \max_{0\le i <width} ( \log(D_{i}))
    \label{eqn:retention_lifespan}
\end{equation}

\subsection{Measurement Density}
Measurement density assesses the importance of measurements in a circuit. A higher measurement count implies the fact that each measurement might be of relatively less importance (e.g., periodic measurement in QEC, or measurement over ancilla qubits), whereas for application with less measurements, the measurement may be of utmost importance. The importance also increases when a measurement accounts for a wider and/or deeper circuit. A good example is the SWAP test (see Figure~\ref{fig:swap_test_res}), where the circuit can be very large but only one measurement is taken to report the similarity. Consequently, this measurement is extremely important to the application. Measurement Density is defined in Equation~\ref{eqn:measurement_density}, where $N_{\text{measurement}}$ is the number of measurements in an application. Since the circuit depth/width can be large and the importance of measurement decays when circuit depth/width keeps on increasing, we add a log to shrink the scale.
\begin{equation}
    \text{Measurement Density} = \frac{\log{(\text{Circuit Depth}\times\text{Circuit Width})}}{N_\text{measurement}}
    \label{eqn:measurement_density}
\end{equation}

\subsection{Entanglement Variance}
Entanglement Variance measures the balance of entanglement across the qubits of a circuit. Circuits with a higher Entanglement Variance indicate that certain qubits are more connected than other qubits (i.e., using more 2-qubit gates such as \texttt{CX} than others). This metric implies that when the circuit is mapped to a NISQ device: (i) less \texttt{SWAP} gates are needed if those hotspot qubits are mapped to the central vertices in the NISQ device topology, such as Qubit-1 in \emph{ibmq\_belem} and Qubit-2 in \emph{ibmq\_yorkton} in Figure~\ref{fig:topo}). A higher entanglement variance implies a higher potential benefit from a good logic-physical qubit-mapping through quantum transpilation. If the entanglement variance is zero, little benefit should be expected from a better transpilation strategy; (ii) Given 2-qubit gate is one of the major sources introducing error, a higher Entanglement Variance implies uneven error introduction among qubits. We define Entanglement Variance in Equation~\ref{eqn:entanglement_variance}, where $G_{q_i}(\text{2-qubits})$ is the number of 2-qubit gates operating on qubit-$i$, and $\overline{{G_{q}}}$ is the average number of 2-qubit gates operating over each qubit. The inclusion of plus 1 is to eliminate log errors with 0 variance. 
\begin{equation}
    \text{Entanglement Variance} = \frac{\log{(\sum_{i=0}^{width}
    (G_{q_i}(\text{2-qubits}) -\overline{{G_{q}}}(\text{2-qubits}))^2+1)}}{\text{Circuit Width}}
    \label{eqn:entanglement_variance}
\end{equation}

\section{Evaluation}

\mm{We first evaluate QASMBench circuits using the proposed metrics and then benchmark a subset of the circuits on real NISQ devices.}

\subsection{Circuit Metrics}
We evaluate QASMBench applications described in Section~2 using the circuit metrics proposed in Section~3.

\subsubsection{Circuit Width} 
\begin{figure}[!t]
\centering
\includegraphics[width=1\columnwidth]{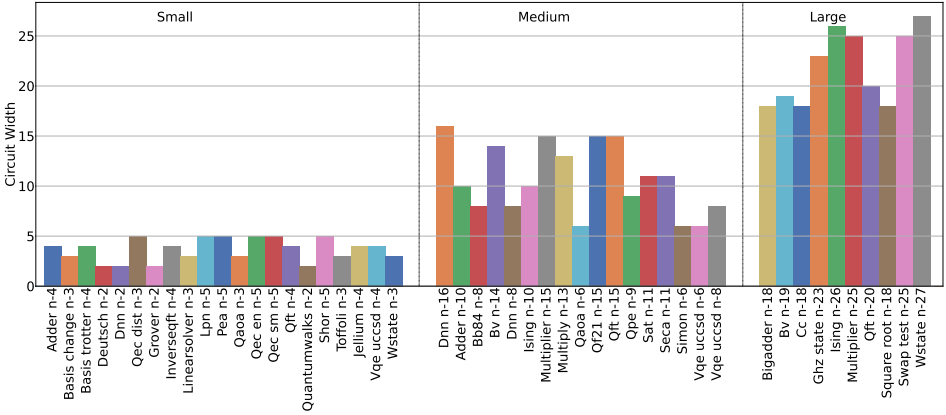} 
\caption{Circuit Width of QASMBench circuits with respect to the small, medium and large categories.}
\label{fig:circuit_width}
\end{figure}

Figure~\ref{fig:circuit_width} illustrates the width of the circuits in QASMBench. Since we partition QASMBench with respect to the number of qubits (1-5 for small-scale, 6-15 for medium-scale, and 15-127 for large-scale), this can be seen in Figure~\ref{fig:circuit_width}.    

\begin{figure}[!t]
\centering
\includegraphics[width=1\columnwidth]{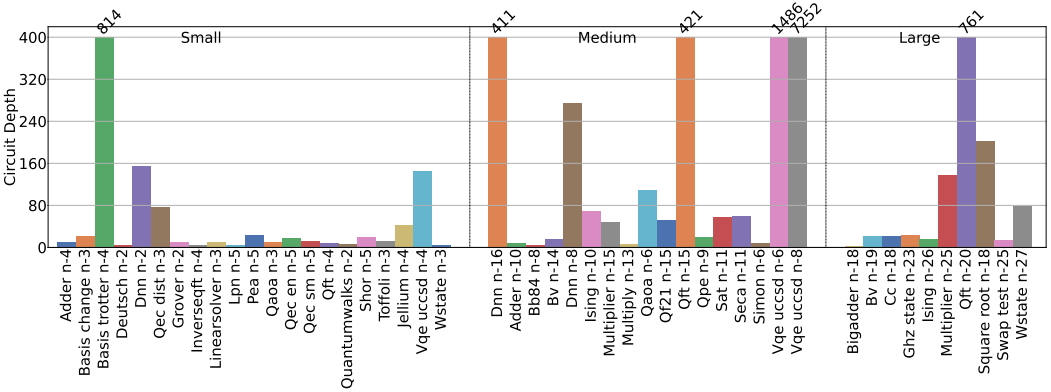} 
\caption{Circuit Depth of QASMBench circuits.}
\label{fig:circuit_depth}
\end{figure}

\subsubsection{Circuit Depth} Figure~\ref{fig:circuit_depth} shows the depth of the benchmark circuits from QASMBench. Apparently, some circuits (e.g., \emph{DNN}, \emph{VQE}, \emph{QFT}) are significantly deeper than other circuits. This figures shows that QASMBench covers a wide range of circuit depth from 11 in the \emph{Linearsolver}\_n-3 to 14,281 in the \emph{QFT}\_n-85, offering a comprehensive circuit database for evaluating NISQ capacity and stability.

\begin{figure}[!t]
\centering
\includegraphics[width=1\columnwidth]{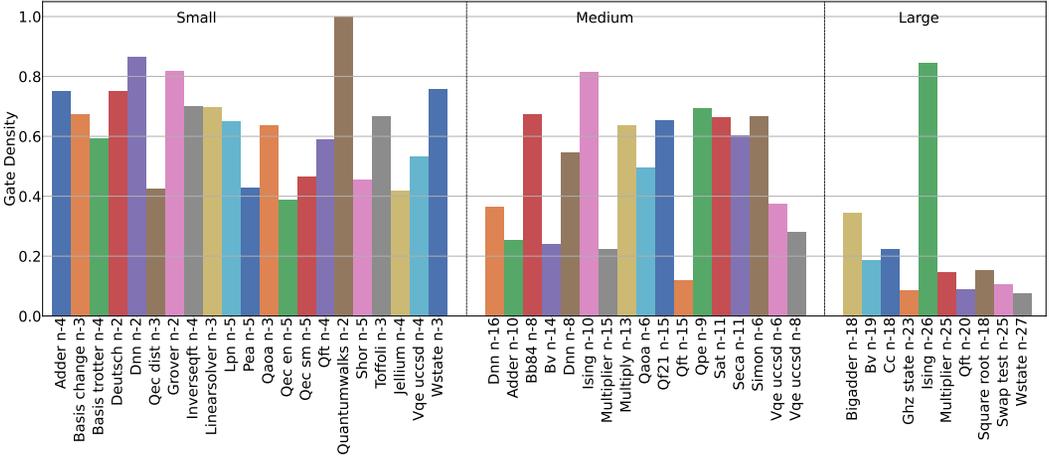} 
\caption{Gate Density of QASMBench circuits.}
\label{fig:gate_density}
\end{figure}

\subsubsection{Gate Density} The plot of Gate Density for QASMBench is visualized in Figure~\ref{fig:gate_density}. The general observation is that, with larger scale circuits, the gate density tends to be more sparse, implying reduced execution occupancy or efficiency. On the contrary, smaller circuits tend to be denser and more carefully designed, so that qubits are less likely to wait for other qubits before evolving onto the next time step. Nevertheless, QASMBench includes densely packed large and medium circuits with \emph{Ising}\_n-26 having a Gate Density factor of 0.8462 for large scale circuits, and \emph{BB84}\_n-8 with a Gate Density of 0.675 in the medium scale category. Note, a low gate density may also imply the amount of internal operation dependency and the potential gain with a better circuit reform.

\begin{figure}[!t]
\centering
\includegraphics[width=1\columnwidth]{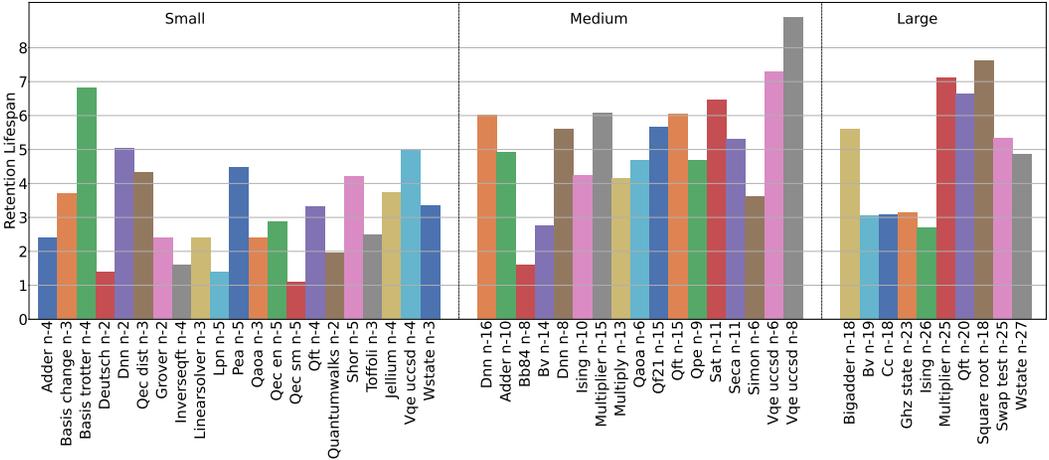} 
\caption{Retention Lifespan of QASMBench Circuits}
\label{fig:retention_lifespan}
\end{figure}

\subsubsection{Retention Lifespan} Retention Lifespan, visualized in Figure \ref{fig:retention_lifespan}, illustrates the highly varying maximum qubit lifespan within the benchmark circuits of QASMBench. Short circuits such as \emph{Deutsch}\_n-2 are required to preserve a quantum state for a relatively short period of time with a retention lifespan of 1.3863, compared to the \emph{Square-Root}\_n-18 circuit with a retention life span of 5.6168. Note, retention lifespan is related to circuit depth, but taking dependency into consideration and is specific to a certain qubit rather than the overall system.

\begin{figure}[!t]
\centering
\includegraphics[width=1\columnwidth]{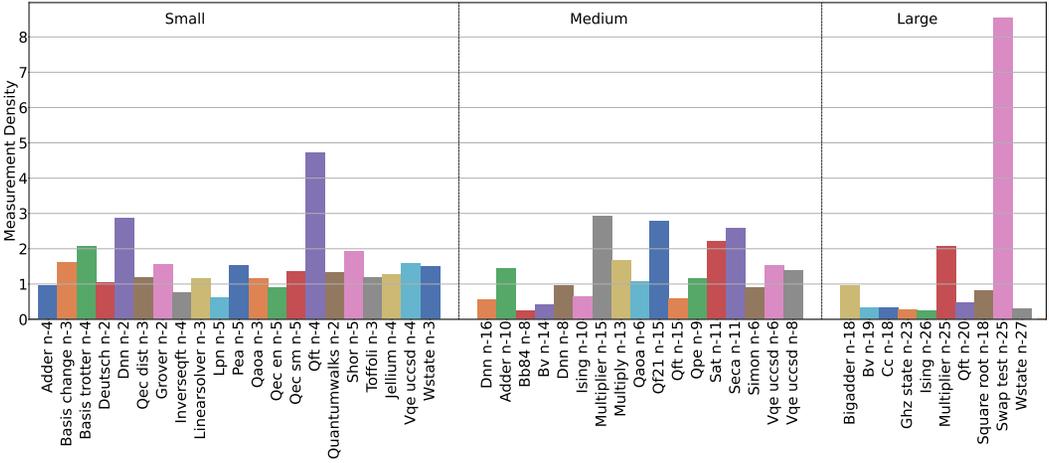} 
\caption{Measurement Density of QASMBench Circuits}
\label{fig:measurement_density}
\end{figure}

\subsubsection{Measurement Density} The measurement density for the QASMBench circuits is illustrated in Figure~\ref{fig:measurement_density}, which measures the importance of measurement operations in a circuit. As can be seen, \emph{SWAP-test}, \emph{multiplier}, \emph{DNN} and \emph{QFT} show significantly higher measurement density than the other circuits. We have already discussed Swap-Test in Section~3.5. \emph{QFT} has a very deep circuit which explains why the only measurement at the end is of more importance. For \emph{multiplier}, only qubits storing the product results are measured at the end. DNN essentially embeds the SWAP-test as a component for the estimation of the cost function.

\begin{figure}[!t]
\centering
\includegraphics[width=1\columnwidth]{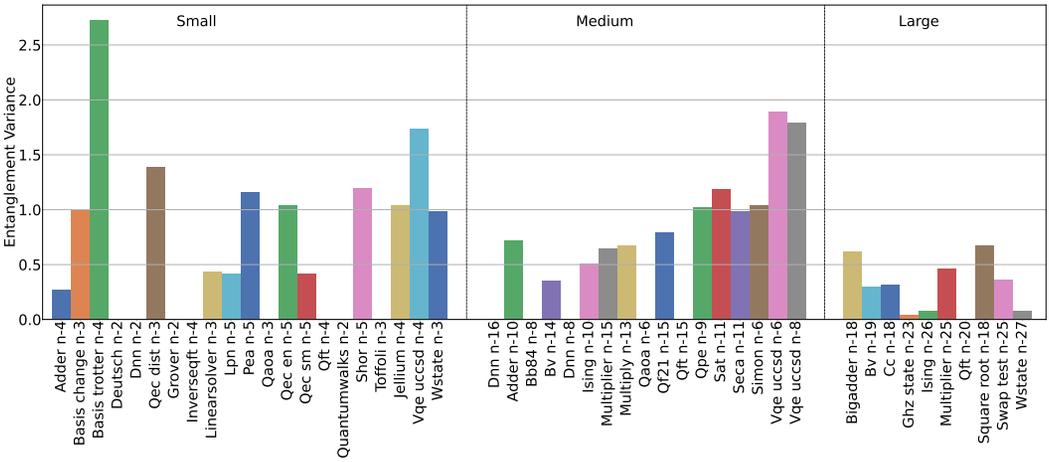} 
\caption{Entanglement Variance of QASMBench Circuits}
\label{fig:entanglement_variance}
\end{figure}

\subsubsection{Entanglement Variance} Figure~\ref{fig:entanglement_variance} shows the entanglement variance for the QASMBench circuits. The figure demonstrates QASMBench's diverse entanglement arrangements, with densely entangled circuits such as Dnn n-2 and Grover n-2, where all entanglement operations entangle the only two qubits available. Therefore, the entanglement is evenly distributed across the circuit. However, compared to Basis trotter n-4, the primary bulk of entanglement is between qubits 1 and 2, whereas qubits 0 and 4 interact substantially less with other qubits. Consequently, when mapping \emph{Basis-trotter} to a NISQ device like \emph{ibmq\_belem}, it involves much less \texttt{SWAP} gates if qubit-1 and 2 are mapped to Q1 and Q3 in Figure~\ref{fig:topo} than the reverse.

\subsection{\mm{Evaluating QASMBench on Real NISQ Devices}}

\mm{We first describe the metric we used for the evaluation. Then, we evaluate applications from the small category of QASMBench on 10 IBM-Q machines. Finally, we compare among devices from three different NISQ hardware vendors.}

\subsubsection{Evaluation Metric}

\mm{We use \textbf{fidelity} to assess the quality of executing a particular quantum algorithm in a specific quantum device. Fidelity here is defined as the distance between the actual quantum state of a NISQ device with the ideal quantum state obtained from a classical simulation. Since NISQ devices are susceptible to noise, the actual quantum state is essentially a mixed state, described by a density matrix. However, quantum observables are not directly measurable on the classical side. Therefore, we reconstruct this mixed state density matrix through state tomography.}  

\mm{Equation~\ref{eqn:fidelity} shows the definition of fidelity we used in the evaluation. Its average value at runtime reflects a quantum machine's ability to induce a particular circuit. }
\begin{equation}
    F(\rho,\sigma) = \left(Tr\sqrt{\sqrt{\rho}\,\sigma\sqrt{\rho}}\right)^2
    \label{eqn:fidelity}
\end{equation}
\mm{where $\rho$ is the density matrix representing the pure state $\ket{\psi_\rho}$ from a classical simulation. $\sigma$ is the density matrix representing the mixed state $\ket{\psi_\sigma}$ from the noisy execution in the NISQ device. Since $\ket{\psi_\rho}$ is a pure state, Equation~\ref{eqn:fidelity} can be further simplified as:}
\begin{equation}
    F(\rho,\sigma) = \bra{\psi_\rho}\sigma\ket{\psi_\rho}    
    \label{eqn:fid_pure}
\end{equation}
\mm{We use Equation~\ref{eqn:fid_pure} to measure the fidelity, which should be the most precise fidelity metric, to the best of our knowledge. However, density matrix tomography is extremely computationally expensive, which requires $3^n$ times' circuit evaluations where $n$ is the number of qubits. Therefore, for a 10-qubit circuit, it demands 59,049 circuit evaluations (without even counting the shots per evaluation). Additionally, these evaluations should be performed consecutively in a short time window and at least within the same calibration period. Otherwise, the state to be reconstructed is not consistent any more. Consequently, such an approach is only feasible for circuits with a small number of qubits. Nevertheless, active research is ongoing in this area to accelerate and make state tomography more scalable through technologies such as neural networks \cite{quek2021adaptive}.}

\subsubsection{IBM-Q Evaluation}

\mm{Due to the excessive cost of density matrix tomography, we evaluate 24 circuits from the small category (see Table~\ref{tab:small}) on 12 IBM-Q machines: \texttt{Washington} (127-qubit, QV64 which means quantum volume is 64), \texttt{Guadalupe} (16-qubit, QV32), \texttt{Toronto} (27-qubit, QV32), \texttt{Brooklyn} (65-qubit, QV32), \texttt{Perth} (7-qubit, QV32), \texttt{Cairo} (27-qubit, QV64), \texttt{Jakarta} (7-qubit, QV16), \texttt{Montreal} (27-qubit, QV128), \texttt{Mumbai} (27-qubit, QV128), \texttt{Auckland} (27-qubit, QV64), \texttt{Lagos} (7-qubit, QV32) and \texttt{Hanoi} (7-qubit, QV32), through PNNL's IBM-Q Hub. We perform density matrix state tomography to calculate the fidelity using Equation~\ref{eqn:fid_pure} for each circuit, with respect to the pure state from the state-vector simulator of Qiskit. This effort comprises 25,000 circuit evaluations in total. We use 8192 shots per evaluation. The Qiskit version is 0.33.0. We use the default optimization level for the Qiskit transpiler.}

\begin{figure}
    \centering
    \includegraphics[width=1\textwidth]{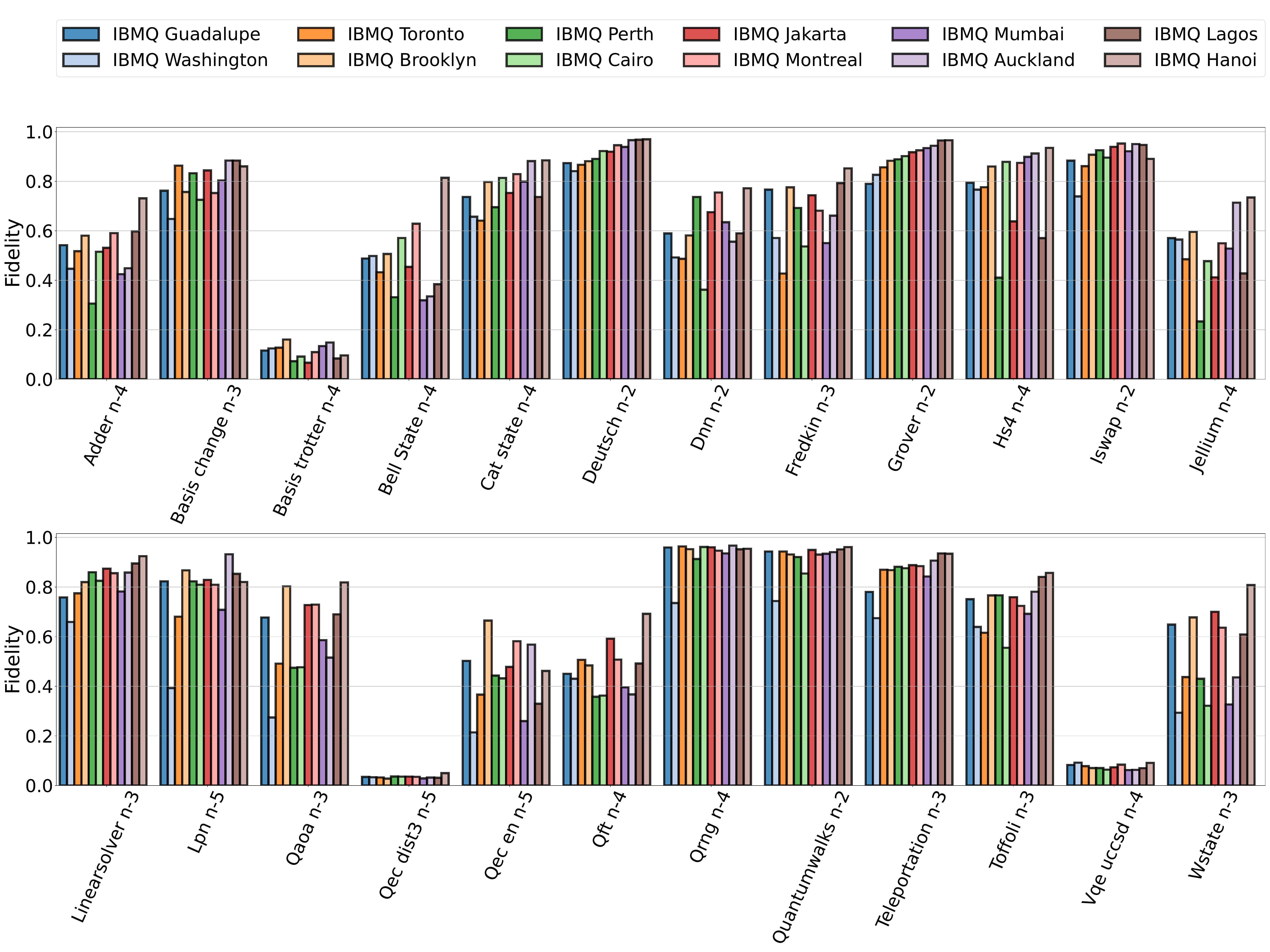}
    \caption{\mm{Evaluations of 24 circuits from QASMBench on 12 IBM-Q machines via density matrix tomography.}}
    \label{fig:ibm_eval}
\end{figure}

\mm{Figure~\ref{fig:ibm_eval} illustrates the results. We observe that: (i) Compared among different benchmark circuits, the fidelity is generally correlated with circuit depth and width. For example, the circuit depths of \texttt{basis\_trotter}, \texttt{qec\_dist3}, and \texttt{vqe\_uccsd} are much larger than the other circuits, with \texttt{qec\_en} and \texttt{qft} stay in the middle. For \texttt{dnn n-2}, although the circuit is relatively deep (268 gates), due to less qubit, the fidelity level is better than \texttt{adder n-4} with merely 23 gates. (ii) Compared among different machines, the fidelity appears to be weakly correlated with the qubit number of the machine and its quantum volume. Particularly, for \texttt{qnn n-2}, despite mapping to the best performing qubits of \texttt{IBM-Q Washington}, the fidelity level is still lower than the 7-qubit \texttt{IBM-Q Hanoi} with a less QV. (iii) In general, the impact from the noise, e.g., the decay noise, is significant. With merely 36 gates in \texttt{qft n-4}, the fidelity for most machines drops below 0.6 except \texttt{Hanoi}.}  

\subsubsection{Evaluations of NISQ Devices}

\begin{figure}
    \centering
    \includegraphics[width=1\textwidth]{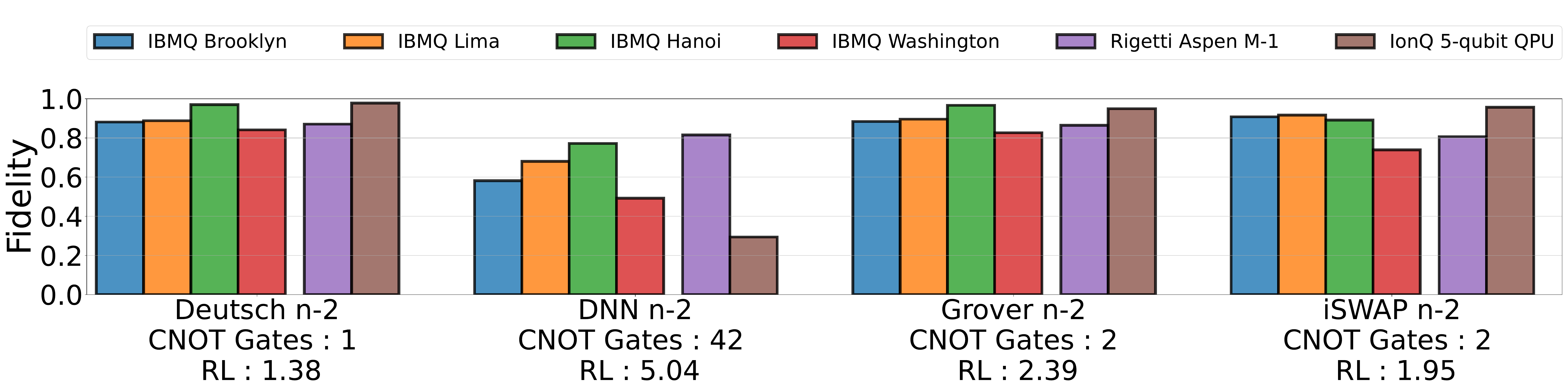}
    \caption{Fidelity evaluation on 4 IBM-Q machines, IonQ's 5-qubit QPU and Rigetti's 80-qubit Aspen M-1 machine. RL refers to the retention lifespan of the circuit (see Equation~\ref{eqn:retention_lifespan}).}
    \label{fig:ionq}
\end{figure}

\mm{We further evaluate 4 circuits (i.e., \texttt{Deutsch}, \texttt{DNN}, \texttt{Grover} and \texttt{iSWAP}) on different NISQ devices. We use the 5-qubit IonQ trapped-ion QPU through Microsoft Azure Quantum \cite{azure}, and the 80-qubit Rigetti Aspen M-1 superconducting machine through Oak Ridge Leadership Computing Facility (OLCF). We compare the fidelity with IBM-Q \texttt{Brooklyn} (65-qubit, QV32), \texttt{Lima} (5-qubit, QV8), \texttt{Hanoi} (27-qubit, QV64), and \texttt{Washington} (127-qubit, QV64). The results are shown in Figure~\ref{fig:ionq}. The modest evaluation is due to resource limitations.}

\mm{Through this evaluation, we can observe that IonQ's trapped-ion system generally obtains higher fidelity than the other devices, for relatively shallow circuits with smaller retention lifespan. However, for a deeper circuit, the performance drops significantly compared to the other superconducting devices. This can be attributed to the lower single-gate fidelity level for the trapped ion processor (i.e., $99.35\%$), compared with the other superconducting devices (e.g., $99.90$\% for IBM-Q \texttt{Washington}). Despite the higher T1 time ($\ge 10^7\mu s$) of IonQ's QPU can preserve quantum information better (with respect to $99.28\mu s$ for IBM-Q \texttt{Washington}), the exponential compounding imperfect gate fidelity leads to an observable reduction in the overall performance.} 

\section{Related Work}

To the best of our knowledge, QASMBench is the first benchmark suite aiming at evaluating NISQ devices using quantum applications from a broad range of domains. Existing works have already proposed to adopt randomly generated benchmarks for evaluating the error rates of different gate operations regarding a particular NISQ device \cite{magesan2011scalable, proctor2019direct, magesan2012characterizing}. Recently, Patel et al. \cite{patel2020experimental} evaluated several IBM-Q machines using seven benchmarks and drew interesting observations regarding the error and execution time. Since the work mainly focused on extracting and validating new insights, only limited applications were included and no new metrics had been proposed. 

Regarding QC metrics, Cross et al. defined the quantum volume protocol which quantified the largest random circuit of equal width and depth that a NISQ device can successfully implement \cite{cross2019validating}, taking gate and measurement error into consideration. 
Quantum Linpack, proposed by Dong and Lin \cite{dong2020random} recently, attempted to develop a random circuit block-encoded matrix (RACBEM) for measuring the QC capability in resolving a random system of linear equations. Benedetti et al. \cite{benedetti2019generative} proposed the qBAS score metric for benchmarking the shallow quantum-classical hybrid systems using synthetic data set. These metrics, despite being comprehensive, are not quite intuitive for understanding. Besides, these metrics all involve certain randomness due to synthetically generated inputs or circuits, which do not necessarily share the properties of real quantum workloads. Given the randomness, the validation requires offline simulation in a classic computer, ultimately limiting the scalability as the simulation cost in a classical computer scales exponentially with increased number of qubits. 

Regarding the general evaluation of NISQ devices, in addition to \cite{patel2020experimental}, LaRose \cite{larose2019overview} compared the four widely used QC programming environments: Forest/pyQuil, Qiskit, Project and QDK/Q\# in terms of library support, hardware, compiler and simulator performance. 
McCaskey et al. \cite{mccaskey2018language} proposed XACC, a programming model that defines a low-level intermediate representation and compiler front-end to provide a unified abstraction for evaluating different quantum hardware and programming environments. Finally, Abhijith et al. \cite{abhijith2018quantum} developed a great hands-on tutorial in explaining the principles of QC using 20 different quantum algorithms, using the 5-qubit IBM-Q machine as the validation platform. However, this work is mainly for education purposes using the small-scale circuits (5 qubits) rather than serving as a QC benchmark suite. \mm{SupermarQ \cite{tomesh2022supermarq} comprises a set of circuits in attempts to benchmark the current term quantum processors. Lubinski et al. evaluated NISQ processors comprehensively according to volumetric benchmarking, and provided evaluations of quantum volume on a diverse set of NISQ superconducting and trapped ion quantum processors \cite{lubinski2021application}.}


\section{Conclusion}

In this paper, we propose a light-weighted, low-level and easy-to-use benchmark suite called QASMBench based on the OpenQASM quantum assembly language. It collects commonly seen quantum algorithms and routines from a variety of domains with distinct properties, serving the need for convenient quantum computing characterization and evaluation purposes for the computer system and architecture communities. Additionally, we propose four novel circuit metrics including gate density, retention lifespan, measurement density, and entanglement variance for assessing the execution efficiency, susceptibility to NISQ error, and potential gain from low-level optimizations. \mm{We also evaluate the fidelity of inducing QASMBench circuits on IBM-Q, IonQ and Rigetti NISQ devices through density matrix state tomography and derive new observations. QASMBench, together with the metrics and evaluations, provides a unique tool for evaluating and characterizing the properties of emerging NISQ devices, estimating the potential of optimization from quantum transpilers, and benchmarking the performance of classical quantum simulators.}

\begin{acks}
This material is based upon work supported by the U.S. Department of Energy, Office of Science, National Quantum Information Science Research Centers, Co-design Center for Quantum Advantage (C2QA) under contract number DE-SC0012704. This research used resources of the Oak Ridge Leadership Computing Facility, which is a DOE Office of Science User Facility supported under Contract DE-AC05-00OR22725. We also acknowledge support from Microsoft's Azure Quantum for providing credits and access to the ion-trap quantum hardware used in this paper. The Pacific Northwest National Laboratory is operated by Battelle for the U.S. Department of Energy under contract DE-AC05-76RL01830.
\end{acks}

\bibliographystyle{ACM-Reference-Format}
\bibliography{qasmben}

\end{document}